\documentclass[aps,prb,twocolumn,superscriptaddress,showpacs]{revtex4}
\usepackage{graphicx}
\usepackage{color}
\usepackage{amssymb}
\usepackage{amsmath}
\newcommand{\FG}[1]{Fig.~\ref{#1}}
\newcommand{\EQ}[1]{Eq.~(\ref{#1})}
\newcommand{\EQs}[1]{Eqs.~(\ref{#1})}
\newcommand{\ea}{{\it et al.}}
\newcommand{\bm}[1]{\mbox{\boldmath$#1$}}
\begin{document}
\title{Native defects in the 
  Co$_2$Ti\bm{Z} (\bm{Z=} Si, Ge, Sn) full Heusler alloys: formation
  and influence on the thermoelectric properties}
\author{Voicu Popescu}
\altaffiliation{E-mail: voicu.popescu@uni-due.de}
\author{Peter Kratzer}
\affiliation{
Faculty of Physics and
Center for Nanointegration (CENIDE),
University of Duisburg-Essen,
Lotharstra{\ss}e\ 1, 47057 Duisburg, Germany}
\author{Sebastian Wimmer}
\author{Hubert Ebert}
\affiliation{
Department Chemie/Physikalische Chemie,
Ludwig Maximilian University Munich, Germany
}

\date{\today}

\begin{abstract}
We have performed first-principles investigations on
the native defects in the half-metallic, ferromagnetic
full Heusler alloys Co$_2$Ti$Z$
($Z$ one of the group IV elements Si, Ge, Sn),
determining their formation energies and how
they influence the transport properties. 
We find that the Co vacancies (Vc),
the Ti$_\text{Sn}$ as well as the
Co$_Z$ or Co$_\text{Ti}$ anti-sites exhibit the 
smallest formation energies.
The most abundant native defects
were modeled as dilute alloys, treated with the 
coherent potential approximation in combination
with the multiple-scattering theory Green function approach. 
The self-consistent potentials determined this way were
used to calculate the residual resistivity via the
Kubo-Greenwood formula and, based on its energy dependence,
the Seebeck coefficient of the systems. The latter is shown
to depend significantly on the type of
defect, leading to variations that are related to
subtle, spin-orbit coupling induced,
changes in the electronic structure above
the half-metallic gap. Two of the systems, Vc$_\text{Co}$ 
and Co$_Z$, are found to exhibit a negative Seebeck coefficient.
This observation, together with their low formation energy,
offers an explanation for the experimentally 
observed negative Seebeck coefficient 
of the Co$_2$Ti$Z$ compounds as being due to 
unintentionally created native defects.
\end{abstract}
\maketitle


\section{Introduction}\label{SecIntro}

Intermetallic ternary compounds often possess
special electronic and magnetic properties that make them interesting
for technological applications. They manifest, however,
a natural tendency towards off-stoichiometry, predictable 
from thermodynamic theory, which creates natural (intrinsic)
doping via vacancies, anti-sites, or swaps.\cite{YZZ17}
Therefore, thorough studies of native defects in these materials 
are required prior to their incorporation in actual devices.
This holds particularly true for ferromagnetic half-metals in view of
possible defect states being introduced in the half-metallic gap.
The present work addresses this issue for Ti-derived half-metallic
Heusler alloys. To motivate this research, we briefly review the use
of half-metals in spintronics.

One class of such materials, under intense scrutiny in the 
last decades, are the Co-based Heusler alloys\cite{KWS83,KFF07,GFP11} 
crystallizing in the cubic $L2_1$ structure. Many of them are 
predicted to be half-metallic, a property describing the particular 
arrangement of the electronic states in which one spin channel is
characterized by metallic conductance while the other one
is semiconducting. While this makes them highly attractive for 
spintronics and spin-caloric transport,
technological achievements are still rare,
owing to the intrinsic complexity of these compounds.
For example, first-principles electronic structure
calculations\cite{GDP02,GOAS06} predicted 
100~\% spin polarization near the Fermi energy in 
Co$_2$MnSi, but it took several more years of research and 
advanced sample preparation to provide experimental evidence 
in either bulk or thin film phases.\cite{YSK+12,JMB+14} 
The theoretically predicted stability of the Co$_2$MnSi/MgO 
interface\cite{HSK09a} soon found experimental validation.\cite{LTM+11}
As a result, several successful implementations
of Heusler-based magnetic tunnel junctions (MTJ),
with potential use as magnetic random access memories (MRAM),
have been reported.\cite{IIT+09,YSK+12,LHT+12,SMSR13}
Consisting of Co$_2$MnSi, Co$_2$FeSi, or Fe$_2$CoSi with 
either MgO or Al-O barriers, they are all characterized by extremely
large, several hundred percent, magneto-resistance ratios.
Other envisioned functionalities of
half-metallic Heusler compounds include
the electrically\cite{CGC+11} and 
thermally driven\cite{CGE+14,GKP14} spin injection,
or the magneto-caloric memory device,\cite{GK15} 
which are yet to be accomplished experimentally.

Apart from Mn and Fe, early transition metals such as titanium offer
further possibilities to tune the half-metallic properties and thus to
open up novel applications. 
Thin films of Co$_2$TiSi were grown on MgO\cite{MSW+11} 
as well as on GaAs substrates.\cite{DH15}
In a previous study, we have proposed  a thermal 
spin injector based on a thin, lattice-matched  barrier 
layer of either Co$_2$TiSi or Co$_2$TiGe
between metallic leads.\cite{GKP14}
Moreover, we could show that, similar to the insulator case, 
the loss of half-metallicity is localized 
within few atomic layers in the proximity of the interface. 
Thus, its role in the filtering of transmission channels
notwithstanding, we find the understanding of 
the {\em bulk} transport properties of particular importance for 
the design of Heusler-based heterostructures.

This aspect appears more critical when considering the example
of the half-metallic Co$_2$Ti$Z$ compounds ($Z=$ Si, Ge, and Sn), 
which made the subject of an extensive and exhaustive study.\cite{BFB+10} 
Ground-state properties,
including stable crystal structure, lattice constants, 
or spin magnetic moments, derived from first-principles calculations 
agree with the experimental findings.\cite{LLBS05,KFF07,SSK10,BFB+10}
On the other hand, theory has failed so far to satisfactorily explain the
transport properties of these systems, in particular
the Seebeck coefficient, whose spin dependence is at the 
core of many spin-caloric proposed applications.
While from the calculated electronic structure at 
and around the Fermi energy one would expect 
a positive Seebeck coefficient in all three Co$_2$Ti$Z$
compounds,\cite{BFB+10,GKP14} 
measurements indicate negative values
throughout the whole temperature range,
reaching $-50$~$\mu$V/K in the 
case of Co$_2$TiSn close to its Curie temperature.\cite{BFB+10,BOG+10}

This proved to be a rather puzzling finding in view of the useful and
robust information that can be extracted from the Seebeck coefficient
in semiconductors. There, a positive (negative) Seebeck 
coefficient can be unambiguously associated with a $p$-($n$-)type 
doping of the semiconductor, a property that led to a rather 
common designation of conduction being either hole- or electron-like,
depending on the sign of $S(T)$. 

Unlike in semiconductors, where its size can be in the 
order of $10^2-10^3$~$\mu$V/K, the Seebeck coefficient of
metallic and half-metallic 
systems is much smaller and thus more sensitive even to 
small variations in the chemical composition of the sample. 
This appears indeed to be the case also for the
Co$_2$Ti$Z$ Heusler alloys, for which small deviations
from the 2:1:1 stoichiometry, deemed to qualify as 
''within the experimental error bars'', 
have been reported.\cite{BFB+10}
 
To what extent the native defects occurring in
the Co$_2$Ti$Z$ systems may influence the thermoelectric properties
is illustrated by the calculated Seebeck coefficient displayed
in \FG{SeebeckCo2TiSiDef}.
This figure, which summarizes the main results we report in this paper,
shows the $S(T)$ curves
corresponding to the Co$_2$TiSi Heusler compound including selected
intrinsic defects at $3$~\% atomic concentration. 
As can be seen, the calculations reveal the occurrence
of a broad range of values, apparently varying between 
hole- and electron-like conduction, depending on the
various off-stoichiometric deviations considered.

\begin{figure}
 \includegraphics[width=0.49\textwidth]{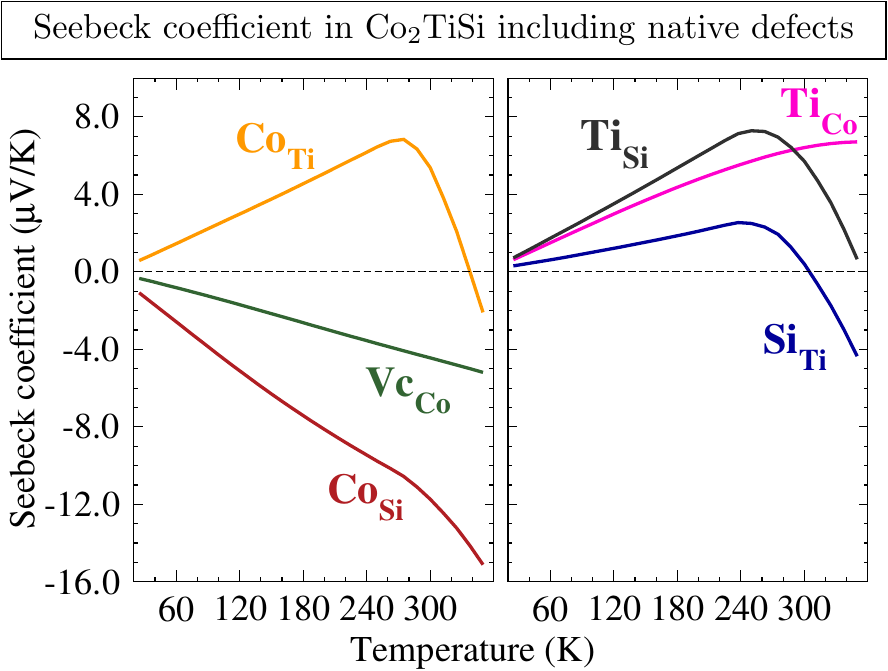}
 \caption{(Color online) Seebeck coefficient calculated for
          several off-stoichiometric native defects in 
          Co$_2$TiSi, modeled as dilute alloys with $3$~\% concentration.
          Note that the same scale is used for $S(T)$ in both panels.}
  \label{SeebeckCo2TiSiDef}
\end{figure}

We investigate here native point defects in
the full Heusler Co$_2$Ti$Z$ alloys, 
including vacancies (Vc) on different sublattices, 
anti-sites, and various sublattice swaps,
with emphasis on their formation energy and their influence on both
the  electronic structure and transport properties of the host
material. Owing to the close-packed character
of the $L2_1$ structure we have deemed more complex defects,
e.g., interstitial site occupation, dumb-bells, or 
clustering effects, as being less likely to occur.
Furthermore, we did not consider the formation and interaction
of defect pairs.
The main focus of our investigations is to
establish a connection between the different features of the
Seebeck coefficient seen in \FG{SeebeckCo2TiSiDef} with 
the defect-induced modifications in the electronic structure.
In particular, we show that the negative $S(T)$ obtained
for Co$_Z$ and Vc$_\text{Co}$ defects
is due to subtle changes of the electronic 
structure that are mediated by spin-orbit coupling. 
Not only are these results in qualitative agreement
with experiment, but they also correspond to defects which
show a small formation energy.

After briefly providing the computational details in 
Section~\ref{SecCompute}, the manuscript comprises two parts
that, although methodologically independent, represent
necessary logical steps in our quest. 
First, we employ pseudopotential-based,
large supercell total energy calculations
to determine the formation energy\cite{VdWN04,ZWZ97,SDM16} for
all possible intrinsic defects. The results, which are
discussed in Section~\ref{SecEFormDef}, also imply
the evaluation of the formation energy by taking
into account the limits of the chemical potentials as obtained from
separate calculations of competing binary and ternary compounds. 
For the second part, we select six different defects, found to 
exhibit formation energies under $\simeq 1$~eV, and 
model them as dilute alloys within the framework of the
coherent potential approximation\cite{Sov67,Tay67,Sov70} 
based on multiple-scattering theory as implemented in the 
full-potential spin-polarized relativistic Korringa-Kohn-Rostoker
method.\cite{EKM11,EBKM16,SPR-KKR} 
A detailed analysis of the various influences these
defects have on the electronic structure of the parent compound is
provided in Section~\ref{SecBSFalloys}. This analysis,
based on the Bloch spectral function, reveals signatures of
several impurity bands appearing in the minority-spin channel,
in the proximity of the half-metallic gap. With spin-orbit
coupling causing majority-minority-band anti-crossings 
right above the Fermi energy, we further evidence the
effect of the different defects on these features, that
may lead, in some cases, to an enhanced conductivity
and a significant reduction of the spin polarization.
The last section presents the results obtained for the
electronic conductivity and the Seebeck coefficient in
the dilute alloys from calculations performed employing
the Kubo-Greenwood formula.\cite{But85,BEV87}
This approach has been already successfully applied
to the study of various transport properties in Heusler
alloys, ranging from longitudinal 
resistivity in Co$_2$Mn(Al,Si)\cite{KTS10}
to the anomalous Hall effect in Co$_2$(Mn,Cr,Co)Al including
native defects.\cite{KDT13}
Applied here in the limit of dilute impurities, we include
in the calculations the so-called vertex corrections,\cite{But85} 
that are extremely important in this regime. 

\section{Details of the calculations}\label{SecCompute}

While our investigations consist of two separate steps, 
they do share the use of self-consistent field (SCF)
calculations. These provide the crystal potential,
determined within density functional theory (DFT),
employing the generalized gradient
approximation (GGA) parameterization for 
the exchange-correlation functional.\cite{PeBu96}
Specific details of the two adopted approaches
are given in the following.

\subsection{Formation energies from large supercells}

The determination of the defect formation energies based on 
DFT is nowadays a standard approach, thoroughly described 
in the literature.\cite{VdWN04,ZWZ97,SDM16}  
It involves calculations of the 
total energies, $E_{\text{tot}}[H]$ and $E_{\text{tot}}[H:D]$,
in two systems: The clean host crystal $H$ and the
perturbed one $H:D$. 
These systems are described by large supercells, constructed
by repeatedly stacking the crystal unit cell along 
the spatial directions. The process of replacing host
atoms by impurities $D$ is regarded as a
particle exchange, mediated by 
external reservoirs associated to each atomic species $\alpha$.
The requirement to describe the equilibrium between the perturbed
host and the reservoirs calls for thermodynamic concepts. In the
following, the thermodynamic equilibrium conditions are worked
out at zero temperature and pressure. Denoting by 
$E_\alpha(\text{bulk})$ the calculated total energy of 
the most stable condensed phase (solid or molecule) of 
element $\alpha$, the chemical potential $\mu_\alpha$
of each $\alpha$-reservoir can be expressed as
\begin{equation}
  \label{MuChemIntro}
  \mu_\alpha=E_\alpha(\text{bulk})+\Delta\mu_\alpha\enspace,
\end{equation}
with $\Delta\mu_\alpha$ the reduced chemical potential. In the present
situation of the perturbed host in equilibrium with the reservoirs, it
is required that
\begin{equation}
  \label{DelMuUp}
  \Delta\mu_\alpha\leq 0 \quad\forall\alpha\enspace.
\end{equation}
In other words, the $E_\alpha(\text{bulk})$ provides 
an upper bound to the chemical potential, corresponding to the 
physical condition of atom $\alpha$ not precipitating into its 
stable phase.

With the notations introduced above, the formation 
energy $E_\text{form}[D]$ of a neutral defect $D$ is obtained
from 
\begin{equation}
  \label{EformDEF}
  \begin{split}
  E_\text{form}[D] = & E_{\text{tot}}[H:D] - 
                       E_{\text{tot}}[H] \\
                    &  - \sum_\alpha N_\alpha E_\alpha(\text{bulk}) - 
                         \sum_\alpha N_\alpha\Delta\mu_\alpha    
  \end{split} \enspace,
\end{equation}
where $N_\alpha$ is the number of atoms, either host or impurity, 
that were added to ($N_\alpha>0$) or removed from ($N_\alpha<0$) 
the supercell upon creating the perturbed system.\cite{VdWN04}
Since the $N_\alpha$'s depend on the chosen configuration
and the total energies are calculated directly, it
follows that, amongst the terms on the rhs of \EQ{EformDEF},
the reduced chemical potentials $\Delta\mu_\alpha$ are
the only independent variables, leading to the compact form
\begin{equation}
  \label{EformShort}
  E_\text{form}[D] = \Delta E(H,D) - 
                    \sum_\alpha N_\alpha\Delta\mu_\alpha  \enspace,
\end{equation}
with
\begin{equation}
  \label{DeltaEform}
  \begin{aligned}
  \Delta E(H,D) = &  E_{\text{tot}}[H:D] - 
                       E_{\text{tot}}[H] \\
                    & \quad - \sum_\alpha N_\alpha E_\alpha(\text{bulk})
  \end{aligned} \enspace.
\end{equation}

The values of the reduced chemical potentials 
are subject to various constraints,
one of which is given in \EQ{DelMuUp} as
an upper bound to $\Delta\mu_{\alpha}$. Additional
bounds arise from the following requirements:
First, thermodynamic equilibrium of the constituents 
with the host crystal must hold. Specific to the 
full Heusler alloys Co$_2$Ti$Z$ investigated here, 
this condition reads 
\begin{equation}
  \label{Co2TiZEquMu}
  2\Delta\mu_\text{Co} + \Delta\mu_\text{Ti} + \Delta\mu_Z = 
            \Delta H(\text{Co}_2\text{Ti}Z) \enspace,
\end{equation}
where $\Delta H(\text{Co}_2\text{Ti}Z)$ is the formation enthalpy of
the Co$_2$Ti$Z$ compound at zero temperature and pressure.
Second, the reduced chemical 
potentials of either host or impurity atoms
have to be in ranges where competing binary or ternary
systems may not form. For a generic compound
with chemical formula Co$_a$Ti$_bZ_c$ and
formation enthalpy $\Delta H(\text{Co}_a\text{Ti}_bZ_c)$
one obtains
\begin{equation}
  \label{MuChemLowBo}
  a\Delta\mu_\text{Co} +   b\Delta\mu_\text{Ti} + c\Delta\mu_Z 
         \leq \Delta H(\text{Co}_a\text{Ti}_bZ_c)\enspace.
\end{equation}

Combining all the constraints derived from the
relations~(\ref{DelMuUp}), (\ref{Co2TiZEquMu}), and 
(\ref{MuChemLowBo}) allows one to determine a stability
range $\{\Delta\mu_\text{Co},\Delta\mu_\text{Ti},\Delta\mu_Z\}$
for which the investigated compound (here the Co$_2$Ti$Z$ Heusler
alloy) may form under equilibrium growth conditions. 
For ternary compounds it is common, using \EQ{Co2TiZEquMu},
to eliminate one of the variables 
(in the following $\Delta\mu_Z$), leaving for
both the stability range and the defect formation
energy $E_\text{form}[X]$ only a two-dimensional 
explicit dependence (in the following 
$\{\Delta\mu_\text{Co},\Delta\mu_\text{Ti}\}$). 

\subsection{Total energy calculations}

The Co$_2$Ti$Z$ ($Z=$ Si, Ge, Sn) compounds investigated here
belong to the class of full Heusler alloys\cite{GFP11} 
of prototype Cu$_2$MnAl, crystallizing in the cubic $L2_1$ 
structure. This crystal structure, shown in \FG{FigStruc}(a), 
has a face-centered-cubic (fcc) primitive cell with four 
inequivalent atomic sites. It can be viewed 
either as four inter-penetrating fcc sublattices, respectively 
occupied by the Co, Ti, Co, and $Z$ atoms
shifted against each other by $(1/4,1/4,1/4)$
lattice constants, or as two inter-penetrating CoTi and 
Co$Z$ zinc-blende structures shifted by $(0,0,1/2)$.

We have considered several off-stoichiometric native
defects: (a) An additional Co, Ti, or $Z$ atom occupying a different
sublattice (in the following denoted as Co$_\text{Ti}$,
Ti$_\text{Co}$, $Z_\text{Co}$ and so on); and (b) Vacancies
on one of the Co, Ti, and $Z$ sites (labeled respectively
as Vc$_\text{Co}$, Vc$_\text{Ti}$, Vc$_Z$). Stoichiometry-preserving
configurations, in which two neighboring atoms from different sublattices
are switching places 
($\text{Co}\leftrightarrow\text{Ti}$, 
$Z\leftrightarrow\text{Ti}$, and $\text{Co}\leftrightarrow{Z}$), 
were also investigated. 
Illustrative examples of the various systems
are depicted in \FG{FigStruc}(a). Note that
a 128 atom supercell (that is, larger than shown in the figure) 
was used in all calculations of the defect formation energy. 
On the other hand, the determination of the competing 
compounds formation enthalpy needed in \EQ{MuChemLowBo}
relied on their respective primitive cells.

\begin{figure}
 \includegraphics[width=0.49\textwidth]{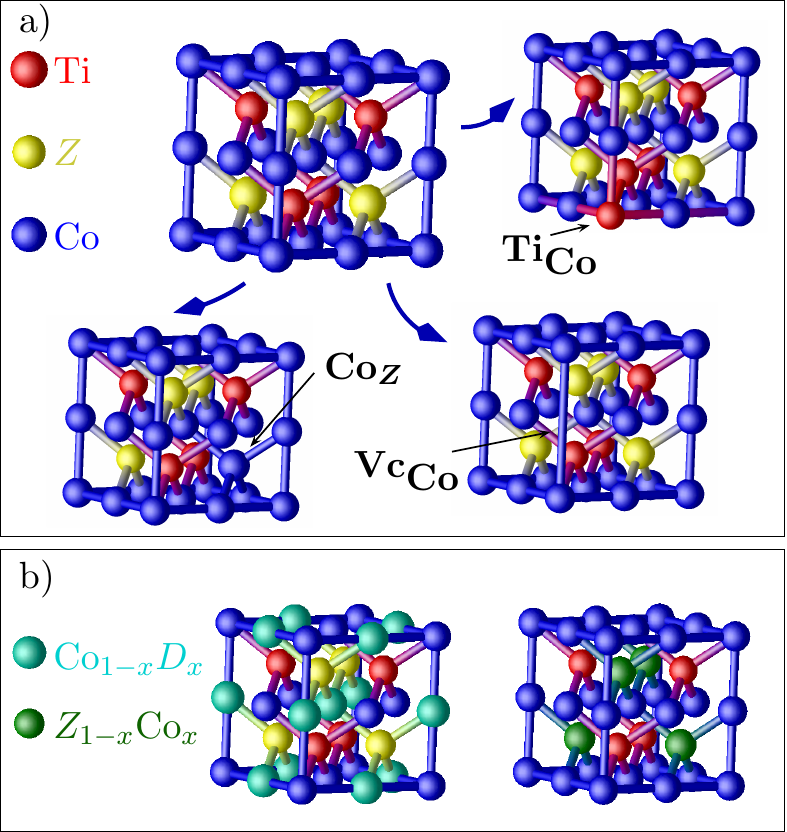}
     \caption{(Color online) (a) Structural model of the
   Co$_2$Ti$Z$ ($Z=$ Si, Ge, Sn) full Heusler 
   $L2_1$ structure and illustrative examples of some of 
   the intrinsic defects investigated in this work. 
   (b) Examples of unit cells used in the alloy modeling 
   of the intrinsic defects in Co$_2$Ti$Z$, accounting only
   for partial disorder: 
   Co(Co$_{1-x}D_x$)Ti$Z$ for a $D$ atom in the Co sublattice 
   ($D_\text{Co}$) and Co$_2$Ti$Z_{1-x}$Co$_x$ 
   for a Co atom in excess on the
   group IV element sublattice (Co$_Z$).}
     \label{FigStruc}
\end{figure}

In all cases, the total energies were calculated for 
spin-polarized systems in the scalar relativistic 
approximation (SRA)
employing the plane-wave pseudopotential method
as implemented in the Quantum Espresso code.\cite{PWSCF}
Wave functions and density have been expanded into plane waves
up to cut-off energies of $40$~Ry and $400$~Ry, respectively.
The neighborhood of atomic centers has been approximated by
self-created ultrasoft pseudopotentials (USPPs),\cite{Vanderbilt:1990}
as described previously in our study of the
Al/Co$_2$Ti$Z$ heterostructures.\cite{GKP14}
A Methfessel-Paxton smearing\cite{MePa89} of 10~mRy
has been applied to the Brillouin zone (BZ) sampling
performed with different Monkhorst-Pack $k$-point grids.\cite{MoPa76}
These were chosen in such a way that they did not include the 
$\Gamma$ point and delivered well-converged total energies and
potentials. In all systems, the lattice constants and 
the internal atomic positions have been 
accurately optimized using Hellmann-Feynman forces to reduce the 
force components below $1$~mRy/$a_0$ 
and the energy changes below $0.1$~mRy.

\subsection{Alloy modeling for the calculation of 
               transport properties}\label{SecCPA}

In a second step, we have determined the 
transport properties of the Co$_2$Ti$Z$ Heusler 
compounds in the presence of the intrinsic defects,
which were modeled as off-stoichiometric dilute alloys
within the CPA.\cite{Sov67,Tay67,Sov70}
As shown schematically in \FG{FigStruc}(b), 
each defect was assigned a mixed occupation of a single
sublattice in the $L2_1$ structure. We illustrate here
examples of Co(Co$_{1-x}$$D_x$)Ti$Z$ and 
Co$_2$Ti($Z_{1-x}$Co$_x$) chemical formulas,
respectively describing the
Co$_2$Ti$Z$:$D_\text{Co}$ ($D=$ Vc, Ti, $Z$)
and Co$_2$Ti$Z$:Co$_Z$ systems. The effect of a
varying defect concentration was accounted for
by considering different $x$ values
($x=0.01$, $0.03$, and $0.05$).

The calculations were based on a 
full potential, spin-polarized
relativistic implementation of the
Korringa-Kohn-Rostoker Green function method 
(FP-SPR-KKR).\cite{EKM11,EBKM16,SPR-KKR}
We obtained the SCF charge density and potentials
by integrating the Green function in the complex energy 
plane over a contour consisting of 36 points and
applying an angular momentum cut-off of
$l_{max}=3$ for the Green function expansion.
We need to emphasize here on the importance of
going beyond the atomic sphere approximation (ASA) when 
methods employing atoms-centered basis functions are
applied to Heusler alloys. 
Even in the case of the closed-packed $L2_1$ structure,
ASA-based electronic structure calculations may fail
to reproduce the FP results.\cite{PCF02,KKW+07}
As pointed out by Picozzi \ea\cite{PCF02}
such discrepancies originate from the
appreciable asphericities that may be 
present in the charge density.\cite{FPvsASAnote}

For all three parent compounds Co$_2$Ti$Z$ 
we calculated the equilibrium lattice constants by
minimizing the total energy. The differences 
between the FP-SPR-KKR determined lattice constants
and those obtained by the scalar-relativistic
plane-wave approach were found to be 
in the range of $\simeq 0.025$~\AA,
with an overall agreement within $1$~\% of the
experimental results.\cite{WZ73,CSP+96,BFB+10}
Including spin-orbit coupling (SOC) has therefore
a very small influence on the computed equilibrium
lattice parameters. We will show, however, that
its role in the thermoelectric properties 
can not be neglected.

The transport properties of the Co$_2$Ti$Z$:$D$ systems
were determined subsequently to the SCF calculations.
More specific, the temperature-dependent
longitudinal Seebeck coefficient 
$S_{ii}(T)$ (with $i=x$, $y$, or $z$ the Cartesian
coordinate) can be obtained from the
diagonal elements of the energy dependent
conductivity tensor $\sigma_{ii}(E)$.\cite{WKE14}
Introducing the transport coefficients
\begin{equation}\label{Lmoms}
  L^{(m)}_{ij} = -\frac 1e
     \int\! \left[
  \frac{\partial}{\partial E} f_0(E,\mu,T)\right]
  \left(E-\mu\right)^m\sigma_{ij}(E)\,dE\enspace,
\end{equation}
where $f_0(E,\mu,T)$ is the Fermi-Dirac distribution function
with chemical potential $\mu$ at energy $E$ for the
temperature $T$, the Seebeck coefficient is given by
\begin{equation}\label{EQSeb}
  S_{ii}(T)  =  -\frac{1}{eT}\,\frac{L^{(1)}_{ii}}{L^{(0)}_{ii}}
\end{equation}
with $e$ the elementary charge.\cite{SebNote}
In our calculations, the central quantity is represented by the
electronic conductivity, obtained on the basis 
of the Kubo-Greenwood formula,\cite{But85,BEV87} appropriately
extended for non-spherical potentials, and including the
important contributions stemming from the
so called vertex corrections.\cite{But85}  
At each energy argument, 
the diagonal elements of $\underline\sigma(E)$ are obtained
through a BZ integral evaluated over a number of
$2\cdot 10^6$ $k$-points. For the energy integrals appearing 
in \EQ{Lmoms} we started by explicitly calculating
$\underline\sigma(E)$ on an equidistant mesh 
of $1$~mRy separation, then refining it to
a $0.01$~mRy resolution by linear interpolation.
The integration boundaries around the chemical potential
were set by the cut-off criterion 
$\partial f_0(E,\mu,T)/\partial E\geq 10^{-3}$
for a fixed electronic temperature $T$.

\section{Defect formation energy results}
\label{SecEFormDef}

We present in this section the results obtained 
by applying the {\em ab initio} thermodynamics concepts
introduced above. After establishing the 
stability ranges by analyzing the competing binary and
ternary compounds, we derive upper and lower boundaries for the 
defect formation energies $E_\text{form}$.
We show that the 
smallest values of $E_\text{form}$ occur for 
vacancies on Co sublattice (Vc$_\text{Co}$)
followed by other defects,
such as the Co anti-sites Co$_\text{Ti}$ and Co$_Z$.

\subsection{Boundaries of the reduced chemical potentials}

The calculated stability domains for each of the
Co$_2$Ti$Z$ full Heusler alloys are displayed in 
\FG{Co2TiZDeltaMu} as two-dimensional representations
in the ($\Delta\mu_{\text Co}$,$\Delta\mu_{\text Ti}$)-plane.
Their construction is based on the numerical evaluation of 
Eqs.~(\ref{Co2TiZEquMu}) and (\ref{MuChemLowBo})
with the corresponding (zero temperature and pressure)
formation enthalpies provided in Table~\ref{DeltaHCompounds}. 
For consistency reasons, these were obtained using 
the same set of self-constructed 
pseudopotentials as employed in the supercell-related 
defect calculations, and were found in 
very good agreement with equivalent data
available in various on-line repositories.\cite{MatProj}
The list of possible competing binary 
phases in Table~\ref{DeltaHCompounds}
is obviously not exhaustive. On the one hand side, we have
only considered compounds known to exist. On the other
hand, rather than exploring all possibilities,
we only focused on those systems
that pose restrictions for the reduced chemical potentials.
In other words, from literature data we could conclude
that certain stoichiometries will not give stricter 
boundaries than the ones already identified. This
procedure appeared to be justified by the overall
good agreement of our data with the previously published
results.

\begin{figure*}
  \centering
  \includegraphics[width=0.96\textwidth]{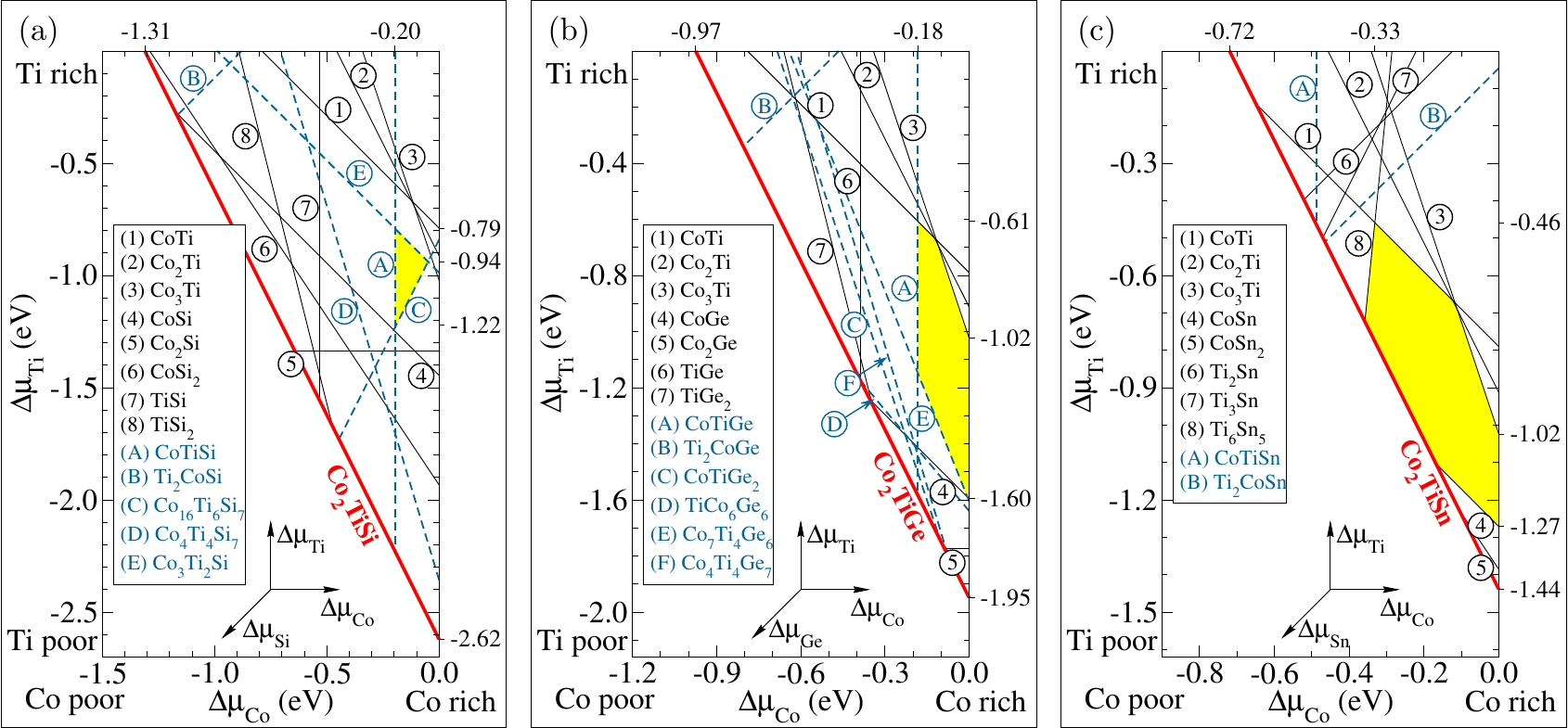}
    \caption{(Color online) 
      The calculated stability regime for the
      full Heusler alloys Co$_2$Ti$Z$ ($Z=$ Si, Ge, Sn)
      in the ($\Delta\mu_{\text Co}$,$\Delta\mu_{\text Ti}$) plane.
      From left to right:
      (a) Co$_2$TiSi, (b) Co$_2$TiGe, and (c) Co$_2$TiSn.
      The red (dark grey) thick line represents the equilibrium condition 
      2$\Delta\mu_{\text Co}$+$\Delta\mu_{\text Ti}$+$\Delta\mu_Z=$
      $\Delta H(\text{Co}_2\text{Ti}Z)$, while the shaded areas mark the
      $\Delta\mu_i$ values ($i=$ Co, Ti, $Z$) for which the 
      Co$_2$Ti$Z$ compound may form under equilibrium growth
      conditions. 
      This is determined by taking into account the
      additional boundaries set to the $\Delta\mu_i$'s
      by the formation of the competing binary or ternary compounds
      accordingly listed in each panel.
      The equilibrium conditions for these phases, 
      corresponding to an equality in \EQ{MuChemLowBo},
      are represented by thin lines appropriately labeled 
      using numbers (for the binaries)
      and letters (for the ternaries). The relative positioning
      of an assigned label with respect to the line it labels
      indicates the respective precipitation range of each
      system.}
  \label{Co2TiZDeltaMu}
\end{figure*}

\begin{table}{\tabcolsep1.2ex
  \begin{tabular}{lcccr}
    Compound & $\Delta H$ & 
    \multicolumn{3}{c}{Lattice constant (\AA)} \\
    (structure) & (eV/f.u.) & QE & KKR & Experiment \\\hline\hline
    Co$_2$TiSi ($L2_1$) & -2.62 & 5.756 & 5.780 & 5.74(0) \\
    Co$_2$TiGe ($L2_1$) & -1.95 & 5.848 & 5.874 & 5.83(1) \\
    Co$_2$TiSn ($L2_1$) & -1.44 & 6.092 & 6.119 & 6.07(3)  \\\hline\hline
  \end{tabular}}
  {\tabcolsep0.2ex
  \begin{tabular}{lrclr}
    Compound & $\Delta H$ & \rule{4ex}{0pt} &  
         Compound & $\Delta H$ \\
    (structure) & (eV/f.u.) & &
    (structure) & (eV/f.u.) \\\hline\hline
   CoTi ($B2$)           &  -0.79 & & TiGe ($B27$)       & -1.18  \\
   Co$_2$Ti ($C15$)      &  -0.91 & & TiGe$_2$ ($C54$)   & -1.26 \\\cline{4-5}
   Co$_3$Ti ($L1_2$)     &  -1.02 & & Ti$_2$Sn ($B8_2$)  & -1.00 \\\cline{1-2}
     CoSi ($B20$)         & -1.18 & & Ti$_3$Sn ($D0_{19}$)  & -1.19 \\
     Co$_2$Si ($C23$)     & -1.29 & & Ti$_6$Sn$_5$ (hP22)  &  -4.34 \\ \cline{4-5}
     CoSi$_2$ ($C1$)      & -1.38 & & Co$_4$Ti$_4$Si$_7$ (tI60) & -11.28 \\
                                     \cline{1-2}
    CoGe ($B20$)          & -0.35 & & Co$_{16}$Ti$_6$Si$_7$ (cF116) & -17.52 \\ 
    Co$_2$Ge ($B8_2$)     & -0.18 & & Co$_3$Ti$_2$Si (hP12) & -3.61 \\ \hline
    CoSn ($B35$)      & -0.17  & & CoTiGe$_2$ (oP48) & -1.85 \\
    CoSn$_2$ ($C16$)  & -0.11  & & Co$_4$Ti$_4$Ge$_7$ (tI60) & -7.49 \\
                                     \cline{1-2}
    TiSi ($B27$)      & -1.55  & & TiCo$_6$Ge$_6$ (hP13) &  -3.50 \\
    TiSi$_2$ ($C54$)  & -1.67  & & Co$_7$Ti$_4$Ge$_6$ (cI34) & -8.48 \\ \hline
   Ti$_2$CoSi ($L2_1$) & -1.74  & & CoTiSi ($C23$/oP12)     &  -2.43 \\
   Ti$_2$CoGe ($L2_1$) & -1.49  & & CoTiGe ($C22$/hP9)     &  -1.77 \\
   Ti$_2$CoSn ($L2_1$) & -0.67  & & CoTiSn ($C1_b$)     &  -0.95 \\\hline\hline
  \end{tabular}}
    \caption{The calculated formation enthalpy $\Delta H$ 
     (in eV/formula unit) for the Co$_2$Ti$Z$ ($Z=$ Si, Ge,Sn)
     full Heusler alloys as well as of various Co-Ti, 
     Co-$Z$, Ti-$Z$, and Co-Ti-$Z$ binary and ternary 
     compounds which compete with the Co$_2$Ti$Z$ formation.
     For the latter we also include the 
     equilibrium lattice constants used in this work
     as obtained either through the 
     plane-wave pseudopotential method in 
     the SRA (results labeled as QE) or the full potential 
     SPR-KKR Green function approach (KKR).
     The calculated values are compared with the
     experimental data.\cite{WZ73,CSP+96}}
    \label{DeltaHCompounds}
\end{table}

The discussion of the results shown in \FG{Co2TiZDeltaMu}
starts with some common features of the 
three panels. The stability triangles represent
$(\Delta\mu_\text{Co},\Delta\mu_\text{Ti},\Delta\mu_Z)$
triplets for which the Co$_2$Ti$Z$ compound may form.
They are defined by the
upper bounds of $\Delta\mu_i$ given in \EQ{DelMuUp}.
As already mentioned, the $\Delta\mu_Z$ variable can be eliminated
using the equilibrium condition (\ref{Co2TiZEquMu}), such that
the $\Delta\mu_Z\leq 0$ constraint reduces to
$2\Delta\mu_\text{Co} + \Delta\mu_\text{Ti} \geq 
\Delta H(\text{Co}_2\text{Ti}Z)$. The equality sign 
is the equivalent of the thick red line in each panel 
of \FG{Co2TiZDeltaMu}. Accounting for 
the formation of competing binary and ternary systems, 
according to \EQ{MuChemLowBo}, further reduces the
allowed $\Delta\mu_i$ values, leading to 
the arbitrarily shaped polygons highlighted by 
the shaded areas. Their boundaries,
displayed as dashed and solid lines and
appropriately labeled by the respective chemical formula
in the figure, are briefly discussed below. 

As can be seen in \FG{Co2TiZDeltaMu}, for the
Co$_2$TiGe and Co$_2$TiSn compounds $\Delta\mu_\text{Co}=0$ 
remains as upper bound, whereas 
$\Delta\mu_\text{Ti}$ is limited above by either 
CoTi, Co$_2$Ti, or Co$_3$Ti. Allowed $\Delta\mu_\text{Ti}$ values must
fulfill the conditions
\begin{equation}
  \label{DMuCoDMuTiUp}
  \begin{aligned}
  \Delta\mu_\text{Ti} \leq \Delta H(\text{Co}_a\text{Ti}) -
  a\Delta\mu_\text{Co}\;(a=1,2,3)\enspace,
  \end{aligned}
\end{equation}
over the various $\Delta\mu_\text{Co}$ intervals.
Comparing our results with those of Chepulskii and
Curtarolo\cite{CC09} who calculated the full series of
Co-Ti alloys, we find a very good agreement 
for all Co$_a$Ti ($a=1\ldots 3$) systems.\cite{CurtaNote}
For CoTi$_2$, on the other hand,
these authors report a formation
enthalpy of $-0.873$~eV/formula unit. It is easy to check,
via \EQ{MuChemLowBo}, 
that the stable CoTi$_2$ bulk phase lies inside the
area already covered by one of the Co$_a$Ti compounds.

The Co$_2$TiSi exhibits the peculiar situation of the
stability range being completely determined by the formation of
ternary compounds: Co$_3$Ti$_2$Si and Co$_{16}$Ti$_6$Si$_7$ set the
upper and lower boundary of $\Delta\mu_\text{Ti}$, while the crossing
of their equilibrium lines fix the maximum $\Delta\mu_\text{Co}$ value
at $-0.05$~eV. On the Co-poor (small $\Delta\mu_\text{Co}$) side,
the lower boundary is set by the CoTiSi compound crystallizing in the
orthorhombic $C23$ structure. 
Analogously, the ternary CoTiGe compound in the hexagonal 
$C22$ structure sets the lower boundary of $\Delta\mu_\text{Co}$ in
the Co$_2$TiGe system. In contrast, CoTiSn, with its ground state the
$C1_b$ (half-Heusler) structure, forms outside the Co$_2$TiSn
stability range, with the Co-poor boundary determined by the
Ti-Sn binaries. As seen in Table~\ref{DeltaHCompounds}, TiSi and TiGe
both crystallize in the $B27$ structure.
A stable phase of TiSn, on the other hand, is not known to exist,
although investigations performed by Colinet \ea\cite{CTF09} on 
a series of Ti-Sn binaries predict a negative formation enthalpy
for TiSn in various structures. 
While our calculations, performed for Ti$_2$Sn, Ti$_3$Sn, and
Ti$_6$Sn$_5$, delivered formation enthalpies 
in close agreement with their results, we chose to set
the stability boundary to that provided by the Ti$_6$Sn$_5$ compound.

In the Co-rich, Ti-poor range (bottom-right corner), the
formation of the Co$_2$TiGe and Co$_2$TiSn competes with that of
the corresponding Co-$Z$ binary compounds. 
In both cases the lower boundary of $\Delta\mu_\text{Ti}$
is set by CoGe or CoSn, leading to the constraint
\begin{equation}
  \label{DMuCoDMuTiDown}
    \Delta\mu_\text{Ti} \geq
    \Delta H(\text{Co}_2\text{Ti}Z) - 
    \Delta H(\text{Co}Z)  -\Delta\mu_\text{Co}\enspace.
\end{equation}

Having completely defined the boundaries for 
$\Delta\mu_\text{Co}$ and $\Delta\mu_\text{Ti}$ we note that
these can also be transferred back to $\Delta\mu_Z$ via 
\EQ{Co2TiZEquMu}. In combination with the observed increase 
of $\Delta H(\text{Co}_2\text{Ti}Z)$ within the $Z=$ Si, Ge, Sn 
series, it becomes obvious, from \FG{Co2TiZDeltaMu},
that $Z$-rich conditions can only be attained 
in the Co$_2$TiSn system.
For $Z=$ Si and Ge, either CoTiSi, CoSi or Co$_7$Ti$_4$Ge$_7$ set an
upper bound for $\Delta\mu_Z$.

We close the discussion on chemical potential
boundaries by analyzing the 
formation enthalpy of the Ti$_2$Co$Z$ systems,
as competing inverse full Heusler alloys.
These compounds are stable through the
whole series of group IV elements $Z$,
with the formation being subject to the condition 
$\Delta\mu_\text{Ti}\geq\Delta\mu_\text{Co}+
[\Delta H(\text{Ti}_2\text{Co}Z)-\Delta
H(\text{Co}_2\text{Ti}Z)]$. According to our results
shown in Table~\ref{DeltaHCompounds}, this condition 
[labeled (B) in \FG{Co2TiZDeltaMu}] 
falls inside the Co$_2$Ti$Z$ stability triangle
for all $Z$ atoms. In all cases, however, it
remains outside the shaded areas
indicating that, while competing precipitation of
Ti$_2$Co$Z$ may occur,
it requires significantly different growth conditions
than the respective stable full Heusler alloy.

\subsection{Formation energies of intrinsic defects 
               in Co\bm{_2}Ti\bm{Z}}

We give in the following a survey of the calculated formation energies
of all the defects considered. The results listed in
Table~\ref{TabEformDefMinMax} give the
lower and upper bounds of the formation energy 
$E_\text{form}[D]$ and the reference energies
$\Delta E(\text{Co}_2\text{Ti}Z,D)$.
The latter, defined by \EQ{DeltaEform},
implicitly contains system-specific information
related to intrinsic mechanisms concerning the
defect formation, such as equilibrium bond length and
electronegativity. The former, obtained
from \EQ{EformShort} by inserting 
the extremum values of the $\sum_\alpha N_\alpha\Delta\mu_\alpha$
term, can be seen as the energy required to exchange particles 
with the reservoirs.\cite{SDM16} Our results can be summarized as follows:

\begin{table*}{\tabcolsep0.8ex
  \centering
  \begin{tabular}{l||rr|rr|rr}\hline
  \rule{2em}{0pt} & \multicolumn{6}{c}{$\Delta E(\text{Co}_2\text{Ti}Z,D)\;\to\;
                            (E^{\min}_{\text{form}},E^{\max}_{\text{form}})$ [eV]}
                     \\
                      \cline{2-7} 
  Defect $D$  & \multicolumn{2}{c|}{Co$_2$TiSi} &
            \multicolumn{2}{c|}{Co$_2$TiGe} & \multicolumn{2}{c}{Co$_2$TiSn}
                    \\
                    \hline\hline
  Vc$_{\text{Co}}$   &   0.29 $\to$ & ( 0.09, 0.24) &
                       0.01 $\to$ & (-0.17, 0.01) & 
                       0.28 $\to$ & (-0.08, 0.28) \\
  Vc$_{\text{Ti}}$   &  2.20 $\to$ & (0.98, 2.15)  &
                       2.16 $\to$ & (0.56, 1.55)  &
                       2.37 $\to$ & (1.10, 1.91)  \\
  Vc$_Z$            &  3.21 $\to$ & (1.63, 2.21)  &
                       2.87 $\to$ & (1.83, 2.52)  &
                       2.96 $\to$ & (2.43, 2.96)  \\
\hline
 Co$_{\text{Ti}}$    &  1.90 $\to$ & (0.88, 1.31)  &
                       1.96 $\to$ & (0.38, 1.54)  &
                       2.16 $\to$ & (0.89, 2.03)  \\
 Co$_Z$             &  1.90 $\to$ & (0.37, 1.10)  &
                       1.37 $\to$ & (0.44, 1.12)  &
                       1.20 $\to$ & (0.78, 1.56)  \\
\hline
 Ti$_{\text{Co}}$    &  1.21 $\to$ & (1.80, 2.23)  & 
                       0.85 $\to$ & (1.27, 2.43)  & 
                       1.00 $\to$ & (1.13, 2.27)  \\
 Ti$_Z$             &  1.50 $\to$ & (0.86, 1.72)  &
                       0.66 $\to$ & (0.29, 1.91)  &
                      -0.30 $\to$ & (-0.16, 0.80) \\
\hline
 $Z$$_{\text{Co}}$   &  2.16 $\to$ & (2.96, 3.69)  & 
                       2.31 $\to$ & (2.56, 3.24)  & 
                       3.18 $\to$ & (2.82, 3.60)  \\
 $Z$$_{\text{Ti}}$   &  0.93 $\to$ & (0.71, 1.57)  &
                       1.91 $\to$ & (0.66, 2.28)  &
                       2.39 $\to$ & (1.29, 2.25)  \\
                       \hline
Ti $\leftrightarrow$ Co  & \multicolumn{2}{c|}{2.57} & 
                           \multicolumn{2}{c|}{2.28} &
                           \multicolumn{2}{c}{2.63} \\
Ti $\leftrightarrow$ $Z$ & \multicolumn{2}{c|}{2.18} &
                           \multicolumn{2}{c|}{2.04} &
                           \multicolumn{2}{c}{1.90} \\
Co $\leftrightarrow$ $Z$ & \multicolumn{2}{c|}{2.97} & 
                           \multicolumn{2}{c|}{2.70} & 
                           \multicolumn{2}{c}{3.21} \\
                           \hline\hline
  \end{tabular}}
  \caption{Formation energy $E_\text{form}$ 
    for  the native defects in 
    Co$_2$Ti$Z$ ($Z$ = Si, Ge, Sn), calculated using
    128-atom supercells within the pseudopotential plane wave method.
    The formation energy entries, derived from \protect\EQ{EformShort},
     are given as an
    interval $(E_\text{form}^\text{min},E_\text{form}^\text{max})$ 
    corresponding to the  
    lower and upper bounds of the chemical potentials $\Delta\mu_i$
    and/or their combinations. Also listed are the reference
    values $\Delta E(\text{Co}_2\text{Ti}Z,D)$ defined by 
    \protect\EQ{DeltaEform}.
    For the stoichiometric defects 'A$\leftrightarrow$B' all these
    quantities are equal.}
  \label{TabEformDefMinMax}            
\end{table*}

i) Vacancies in various sublattices (Vc$_\text{Co}$, Vc$_\text{Ti}$,
and Vc$_Z$): Those appearing in the Co sublattice are found to have
the smallest formation energy, its values even turning negative
for Co-poor conditions in Co$_2$TiGe and Co$_2$TiSn.
The Vc$_\text{Ti}$ and Vc$_Z$ point defects have a significantly
larger formation energy, whereby, in each system, 
the most stable sublattice is the one consisting of $Z$-atoms.
A similar trend was also observed in Co$_2$MnSi by
H\"ulsen \ea\cite{HSK09b}, albeit with a somehow larger value,
$\simeq 1$~eV,
for $\Delta E[\text{Co}_2\text{Ti}Z,\text{Vc}_\text{Co}]$,
which represents the upper bound ($\Delta\mu_\text{Co}=0$) of 
$E_\text{form}[\text{Vc}_\text{Co}]$.
Experiments on Co$_2$Mn$Z$, on the
other hand,\cite{KKFH09} while confirming vacancy concentrations 
as high as $2$~\%, suggest a rather
random distribution over the lattice sites.
An ever increasing Vc$_\text{Co}$ concentration was found
to lead to an $L2_1$$\to$$C1_b$ ordering transition 
in Ni$_{2-x}$MnSb Heusler alloys.\cite{NTM+15}
A similar scenario may also occur in Co$_2$TiSi$\to$CoTiSi($C23$)
and Co$_2$TiGe$\to$CoTiGe($C22$), where the compounds in 
the 1:1:1 stoichiometry set the lower bound for $\Delta\mu_\text{Co}$
and thus correspond to small (negative for $Z=$Ge) defect formation
energies for the Vc$_\text{Co}$ vacancies. 
An apparently different situation is found in Co$_2$TiSn, for which we
found the CoTiSn ground state to be the $C1_b$ structure but lying 
outside the Co$_2$TiSn stability range. Investigations performed 
by Nobata \ea\cite{NNK+02} found, on the other hand, 
that certain growth conditions favor the formation of the 
the Co$_{1.50}$TiSn with a half-filled vacancy sublattice
instead of the stoichiometric compound CoTiSn. This way, the 
obtained negative formation energy of Co$_2$TiSn:Vc$_\text{Co}$ is
consistent with the experimental findings. We do emphasize, 
however, that all the calculated defect formation energies correspond
to the limiting case of zero pressure and temperature.

ii) Co anti-sites (Co$_\text{Ti}$ and Co$_Z$):
Following the sequence Si$\to$Ge$\to$Sn,
opposite trends can be recognized in the
two reference energies 
$\Delta E(\text{Co}_2\text{Ti}Z,\text{Co}_\text{Ti})$
and 
$\Delta E(\text{Co}_2\text{Ti}Z,\text{Co}_Z)$
that correspond to the two Co anti-sites.
While the former increases 
with the atomic number of the $Z$-atom,
and thus with the lattice constant of the
Co$_2$Ti$Z$ compound, the latter decreases.
Accounting, however, for the allowed variations
in the reduced chemical potentials,
which enter as $(-\Delta\mu_\text{Co} + \Delta\mu_{\text{Ti}/Z})$,
the two formation energies 
$E_\text{form}[\text{Co}_\text{Ti}]$ and
$E_\text{form}[\text{Co}_Z]$ 
are found to be of comparable size for a given 
system Co$_2$Ti$Z$. This indicates that both
anti-sites may appear with roughly equal probability during the sample
preparation. 

iii) Ti anti-sites (Ti$_\text{Co}$ and Ti$_Z$):
Similar to the previous case, the 
reference energy of the Ti$_Z$ anti-site
exhibits a strong dependence on the $Z$-atom,
but with an even a more significant drop
for the heavier elements. As
seen in Table~\ref{TabEformDefMinMax}, 
$\Delta E(\text{Co}_2\text{TiSn},\text{Ti}_\text{Sn})$
even becomes negative 
One can conclude that, when Ti is in excess, a preference for
the Ti$_Z$ anti-site should be observed,
at the expense of Ti$_\text{Co}$.
This is quite an opposite trend as the one seen
in Co$_2$MnSi, where the Mn$_\text{Co}$ anti-site
was found to have a much smaller reference
energy than Mn$_\text{Si}$.\cite{HSK09b}
As a general observation, we emphasize here
the fact that the lower/upper boundary for an
anti-site $A_B$ becomes the upper/lower boundary with
opposite sign for its $B_A$ counterpart.
Comparing now the two 'paired' anti-sites
Ti$_\text{Co}$ and Co$_\text{Ti}$, we note
again that the allowed $\Delta\mu_\alpha$ 
intervals significantly
influence the formation energy results. 
Indeed, whereas 
$\Delta E(\text{Co}_2\text{Ti}Z,\text{Ti}_\text{Co})$
is about half the size of
$\Delta E(\text{Co}_2\text{Ti}Z,\text{Co}_\text{Ti})$,
the corresponding $E_\text{form}$ values have a
rather broad overlapping interval, with a lower
boundary for the Co$_\text{Ti}$ anti-site.

iv) $Z$ anti-sites ($Z_\text{Co}$ and $Z_\text{Ti}$):
As could be anticipated from the Vc$_Z$ case, the
sublattice formed by the group IV elements appears
to be the most stable. The $Z_\text{Co}$ anti-sites 
exhibit by far the largest formation energies and
are therefore the least likely to occur. The
formation energies obtained for the $Z_\text{Ti}$ 
defects are somewhat smaller, but still larger
than their $\text{Ti}_Z$ correspondents. 
We further note that the 
$\Delta E(\text{Co}_2\text{Ti}Z,Z_\text{Ti})$
values increase along the Si$\to$Ge$\to$Sn series,
in contrast to the trend observed for 
the $\text{Ti}_Z$ anti-sites.

v) swaps (Ti $\leftrightarrow$ Co,Ti $\leftrightarrow$ $Z$,
and Co $\leftrightarrow$ $Z$):
The calculations performed by Picozzi \ea\cite{PCF04}
for the Co$_2$MnSi full Heusler alloy found an
interesting feature in the formation energy 
of the Mn$\leftrightarrow$Co swap: its value closely
matches the sum of the reference energies of 
the Mn$_\text{Co}$ and Co$_\text{Mn}$ anti-sites.
Our results do not reproduce this behavior in
any of the Co$_2$Ti$Z$ compounds, regardless of
the considered swaps. A possible reason may be
the large differences between the corresponding
reference energies of the paired anti-sites, which
could efficiently curtail the swapping tendencies.
As Table~\ref{TabEformDefMinMax} shows, 
we find very large formation energies
for all investigated swaps. 
These findings appear to be in contradiction with
the experimental results obtained for
the Co$_2$TiSn Heusler alloy,\cite{KKW+07}
for which the authors report 
NMR data indicative of the crystal structure 
being partially $D0_3$-like, consistent
with a Ti$\leftrightarrow$Co swap.

To conclude, we found several defects in Co$_2$Ti$Z$ to have quite a
low formation energy, whereas others, in particular the swaps
and those related to the $Z$-atom sublattice, should
occur with reduced probability.
Some of the more likely 
defects have been selected for subsequent
investigations concerning the effect they have
on the electronic structure
and the implications for the transport properties.
These results form the subject of the next sections.

\section{Defects-triggered changes in the
electronic structure}\label{SecBSFalloys}

We present in this section the results obtained employing the
FP-SPR-KKR Green function method, modeling the intrinsic
defects as dilute alloys. We focus on those systems which
were found to have a small formation energy and follow
the changes in the electronic structure brought about
by the defects, emphasizing on specific aspects relevant 
for the transport properties. In doing so, we use the  
Bloch spectral function (BSF),\cite{FS80} 
expressed as the Fourier transform of the configurationally 
averaged Green function
\begin{equation}\label{BSFdefine}
  \begin{aligned}
A(\vec k,E) = & -\frac{1}{\pi N}\text{Im}\sum_{m,n}^N 
                  e^{i\vec k({\vec R}_m  -{\vec R}_n)}\\
              & \quad\int\!d^3 r\langle G(\vec r + {\vec R}_m,
                         \vec r + {\vec R}_n;E)\rangle\enspace,
  \end{aligned}
\end{equation}
performed over $N$ atomic sites with position vectors
${\vec R}_m$ and ${\vec R}_n$ in the crystal lattice. 
This quantity, which can be seen as a $\vec k$-resolved density of states (DOS),
is well-defined also for ordered systems. In this case 
the configurational average, symbolized by $\langle\ldots\rangle$,
drops out and the evaluation takes place for a complex
energy argument $E+i\varepsilon$ with $\varepsilon\to0$,
providing the ordinary dispersion relation $E_{\vec k}$.

Taking the Co$_2$TiSi full Heusler alloy as an 
illustrative example,
we present the BSFs for a series of dilute alloys
modeling the Co$_2$TiSi:$D$ defects. Calculated
on a fully relativistic level, the BSFs are subsequently
projected on their spin components and 
compared with the standard band structure of the parent compound.
This allows us to identify signatures of the defect-induced
minority-spin impurity bands in the proximity of 
the half-metallic gap. We will also show that,
owing to the SOC, a majority-spin d-band that crosses 
the Fermi energy $E_F$ gains minority-spin character.
This rather ubiquitous effect is larger
for the Co-related defects 
Vc$_\text{Co}$ and Co$_\text{Si}$,
significantly reducing the spin polarization near $E_F$.

\subsection{Results for Co\bm{_2}TiSi bulk}

The spin-polarized relativistic electronic structure for
the ordered compound Co$_2$TiSi is displayed in \FG{Co2TiSiBSTR},
with panel (a) depicting the calculated dispersion relation along 
several high symmetry directions in the fcc BZ
and panel (b) the corresponding spin-resolved BSF 
obtained for an imaginary part of
the energy $\varepsilon=0.005$~Ry 
along the $K-\Gamma-X$ path of the BZ.
The spin-resolved results of \FG{Co2TiSiBSTR}(b),
consistent with previously reported 
band structure calculations on a
scalar relativistic level,\cite{BFB+10,SSK10} 
demonstrate the half-metallic character of this system.
The Fermi energy $E_F$, taken here as the energy reference,
is seen to lie close to the upper edge 
of the band gap appearing in the minority-spin channel.

\begin{figure}
  \centering
  \includegraphics[width=0.48\textwidth]{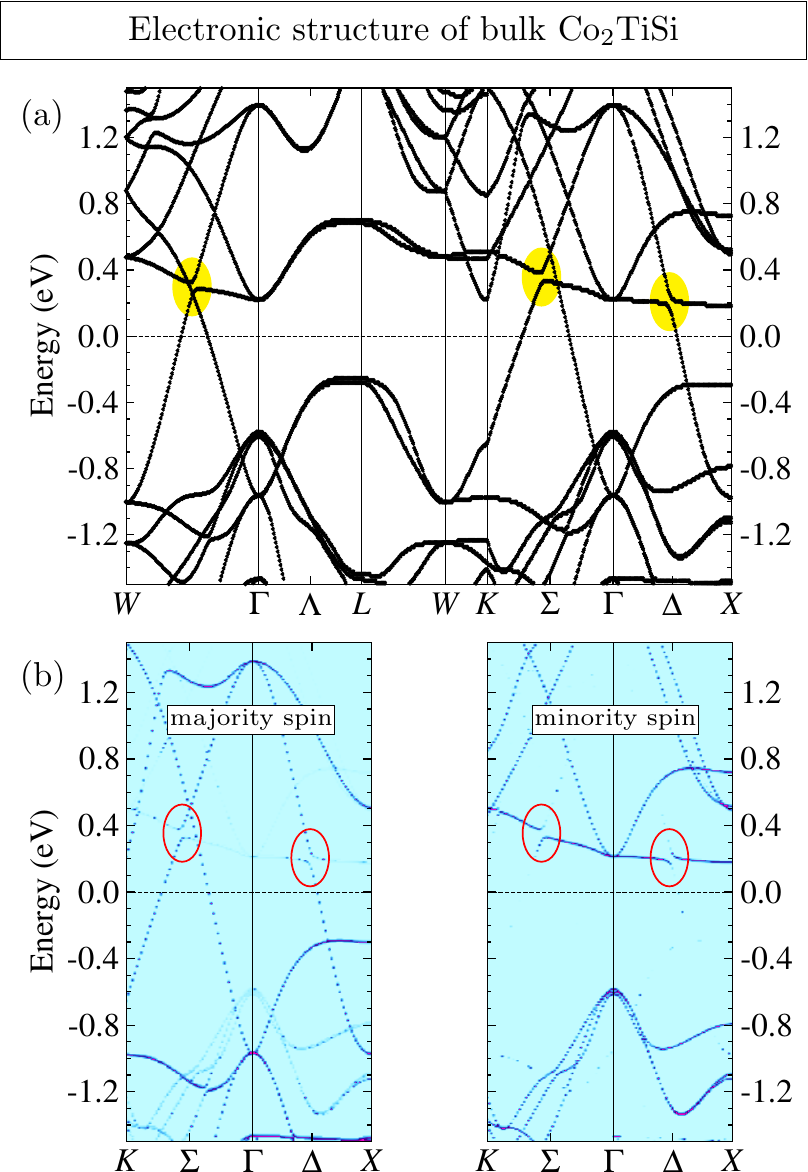}
    \caption{(Color online)
    (a) Spin-polarized relativistic dispersion relation $E_{\vec k}$
    and (b) spin-resolved BSFs along selected high
    symmetry directions in the fcc BZ, calculated for the
    Co$_2$TiSi full Heusler alloy in the $L2_1$ structure
    obtained using the FP-SPR-KKR method. 
    The energy is given relative to the Fermi energy $E_F$
    of the system.
    The highlighted areas located $0.2-0.4$~eV above $E_F$
    along the $W-\Gamma$, $K-\Gamma$, and $\Gamma-X$
    directions point to anti-crossings between 
    a Co majority-d and a Co minority-d band, which
    occur as a result of spin-orbit coupling.}
  \label{Co2TiSiBSTR}
\end{figure}

Of particular interest for the transport properties, in general, 
and for the Seebeck coefficient in particular, 
are the states located in the proximity of $E_F$, 
according to \EQ{Lmoms}. As seen in \FG{Co2TiSiBSTR}(a), 
there are only two bands that cross the Fermi energy.
Starting from $1.4$~eV at the $\Gamma$-point, one band goes
down in energy and gets below $E_F$ at the $X$- and $W$-points.
It thus leads to a Fermi surface enclosing the
$\Gamma$-point and forming pockets at the BZ edge. 
The second band rises in energy from $-1.0$~eV at
the $\Gamma$-point and forms
a Fermi surface connecting to adjacent BZs.
Stemming from the Co d-orbitals, these two bands possess a
dominant majority spin character, as revealed by
the BSF shown in \FG{Co2TiSiBSTR}(b), and 
are expected to dominate the transport properties in 
these compounds. 
A much flatter Co d-band 
is found slightly above the Fermi energy, forming the
minority-spin conduction band. Its minimum lies at
the $X$-point of the BZ, such that the half-metallic gap,
as seen on the right side of \FG{Co2TiSiBSTR}(b), is
indirect.
We note here that, although 
the isoelectronic compounds Co$_2$TiGe and Co$_2$TiSn
exhibit qualitatively similar results, 
one does find quantitative variations in both
the width of the half-metallic gap and the
position of the Fermi energy inside it.
For the former, the FP-SPR-KKR calculations
give $0.76$, $0.57$, and $0.48$~eV, respectively,
for $Z=$ Si, Ge, and Sn. Measured relative to $E_F$, 
the minority-spin conduction band minimum,
on the other hand, was found at $0.18$, $0.21$, and $0.26$~eV.

Highlighted in \FG{Co2TiSiBSTR} are several areas along the
$W-\Gamma$, $K-\Gamma$, and $\Gamma-X$ directions where
this minority-spin d-band intersects the two 
majority-spin d-bands, in a range of $0.2-0.4$~eV above $E_F$.
Qualitative differences in these intersections are
easily recognizable: The falling majority-spin band
has the same parity (and
symmetry) as the minority-spin band, which leads to an 
{\em anti-crossing} between the two bands. 
In turn, the other majority-spin band rising from
the $\Gamma$-point crosses the minority-spin band without 
coupling to it, as seen along the $W-\Gamma$ and 
$K-\Gamma$ directions.

The repelling of the two bands of different 
spin character is caused by SOC. By virtue of
a mechanism suggested by Mavropoulos \ea\cite{MSZ+04} when discussing the 
effect of spin-orbit coupling on the band gap of half metals,
the coupling of the two bands causes a strong spin mixing.
Regarding the SOC as a perturbation to the spin-dependent
crystal Hamiltonian with eigenvalues $E^\uparrow(\vec k)$ and
$E^\downarrow(\vec k)$, the potential terms causing the 
spins to flip are proportional, in leading order, to
$1/[E^\uparrow(\vec k)-E^\downarrow(\vec k)]$. It becomes
apparent that the spin mixing will exhibit a maximum where the 
unperturbed spin-up and spin-down bands would cross.
Indeed, as evidenced in \FG{Co2TiSiBSTR}(b), 
the BSF of a given spin character (here obtained within a Dirac
formalism) is an ''image'' of 
the opposite spin over a broad range around the
anti-crossing points. This effectively leads to a reduced 
spin polarization especially at the top edge
of the half-metallic gap,\cite{MSZ+04} since
the states in its vicinity no longer posses a
unique spin character. 
We note that a similar imaging effect
occurs below the Fermi energy around the $\Gamma$-point. 
As can be seen in \FG{Co2TiSiBSTR}(b), 
one finds a set of valence bands that exhibit
finite amplitudes of the BSF in both spin channels.
The reason for spin mixing 
in this case is related to the 
intra-atomic SOC, with the states originating 
from d-orbitals with large magnetic quantum 
numbers, $\pm5/2$ and $\pm3/2$.

\subsection{Spectral function of dilute alloys}

We have selected, from the list of all possible intrinsic defects
investigated above, a total of six systems, exhibiting low formation
energies. According to the results presented in 
Table~\ref{TabEformDefMinMax}, these are: 
Vacancies on Co sublattice Vc$_\text{Co}$; the Co anti-sites 
Co$_Z$ and Co$_\text{Ti}$; the Ti anti-sites Ti$_\text{Co}$
and Ti$_Z$; and the Si anti-site $Z_\text{Ti}$. 
We employed a dilute alloy modeling for all these systems, 
as described in Section~\ref{SecCPA}, by considering a defect of
type $D_A$ as a single-site effective medium $(A_{1-x}D_x)$
while keeping the other sublattices unchanged.
Without loss of generality, 
the discussion below is restricted 
to the Co$_2$TiSi:$D$ systems ($Z=$ Si) with $x=0.03$,
corresponding to a single defect in the 128-atom 
supercell used in the formation energy calculations. 

Our primary interest is to investigate the changes in the
electronic structure occurring in the proximity of the
Fermi energy. As shown above, only two
bands do cross $E_F$, both having a dominant 
majority-spin character. For this spin channel
we found that, while the BSFs associated with these 
bands may differ quantitatively from one system to another,
no qualitative deviations occur, 
irrespective of the group IV element $Z$ and the
type of defect. 
We shall therefore focus in the following on the minority-spin 
BSFs in a $\simeq 2$~eV energy window around $E_F$,
with emphasis on two aspects: i) the defect-induced states
evolving in impurity bands (IB) near the half-metallic gap;
and ii) the so far unexplored effect of the defects
on the SOC-induced band anti-crossings observed for the ordered
compound. 

The minority-spin BSFs for the off-stoichiometric Co$_2$TiSi:$D$
systems are shown in \FG{Co2TiSiDefBSFdn}. Note that, while
choosing the same $K-\Gamma-X$ direction in the BZ as for the 
Co$_2$TiSi BSFs in \FG{Co2TiSiBSTR}(b), for clarity purposes
the energy interval around the Fermi energy $E_F$
has been reduced. Moreover, since we are dealing now with an
alloy system, \EQ{BSFdefine} is evaluated for real energy
arguments and the broadening of the alloy bands 
is due to the intrinsic disorder of the CPA effective medium. 
We first analyze, on the basis of \FG{Co2TiSiDefBSFdn},
the defect-related states appearing in the investigated systems,
appropriately labeled from (1) to (5) in each panel.

\begin{figure*}
  \centering
  \includegraphics[width=\textwidth,clip]{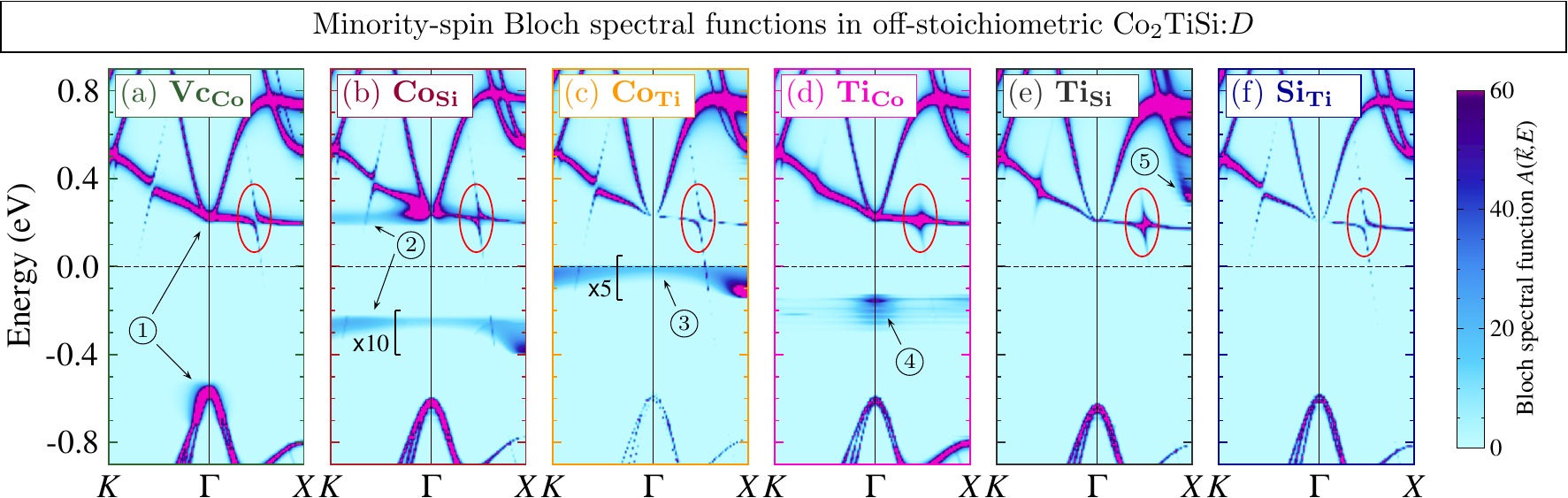}
    \caption{(Color online) Minority-spin BSFs along the
    $K-\Gamma-X$ direction in the fcc BZ, calculated for
    several off-stoichiometric 
    native defects in Co$_2$TiSi modeled as dilute alloys
    with $3$~at.\% defect composition. 
    The energy scale is given relative to the 
    Fermi energy of each system.}
  \label{Co2TiSiDefBSFdn}
\end{figure*}

(1) Vacancies in the Co-sublattice, \FG{Co2TiSiDefBSFdn}(a),
induce minority-spin 
states at the edges of the half-metallic gap,
thus reducing its width and,
consequently, the size of the spin gap. The
latter is defined as the minimum energy 
required to flip the electron spin. The
side-by-side comparison of the different panels of 
\FG{Co2TiSiDefBSFdn} reveals that
the highest minority-spin valence band maximum
is obtained for Co$_2$TiSi:Vc$_\text{Co}$. Actually,
within a broad range around the $\Gamma$-point
the minority-spin BSF of this system is characterized
by a much stronger intensity as compared to the other 
alloys. These results are consistent with those obtained by 
\"{O}zdo\u{g}an \ea\cite{OSG07} for various Co-based
Heusler alloys from the series  Co$_2$(Mn,Cr)(Al,Si).
Similar to these systems, we find that,
although the presence of Co-vacancies reduces the 
total magnetization (e.g., from $2.00$ to $1.90$~$\mu_B$ 
per formula unit in 
the case of $3$~\% vacancy concentration in Co$_2$TiSi), the
half-metallicity is not destroyed. Because of the 
large value of the spin gap, we expect the half-metallicity to
be quite robust against Co-vacancy formation,
at reasonably low concentrations, in all Co$_2$Ti$Z$ compounds.

(2) The Co$_\text{Si}$ anti-site, \FG{Co2TiSiDefBSFdn}(b),
creates two minority-spin IBs near the Fermi energy. 
These are distributed rather symmetrically below and above 
$E_F$, centered around $-0.3$ and $0.2$~eV, respectively. 
Thus, the lower band falls inside the half-metallic
gap, whereas the higher one overlaps with the 
edge of the minority-spin conduction band. In spite 
of the IB present in the gap, the Co$_2$TiSi:Co$_\text{Si}$ also
remains half-metallic. 
As further indicated in \FG{Co2TiSiDefBSFdn}(b), 
the BSF has been multiplied by a factor $10$ in the
energy interval $[-0.4,-0.2]$~eV for all $\vec k$-vectors, a value
that represents the relative weight of the DOS associated to
the high and low IBs. 
Although running rather parallel to each other, these two 
bands differ qualitatively in their dependence on $\vec k$. 
While the upper IB has a large weight
near the $\Gamma$-point, enhancing and significantly broadening 
the BSF, the low one exhibits a maximum near the $X$-point,
at the edges of the BZ. 

(3) The Co$_\text{Ti}$ anti-site, \FG{Co2TiSiDefBSFdn}(c), 
creates a single IB in the energy region of interest.
Located inside the half-metallic gap, this band is
nearly touching the Fermi level, such that
the Co$_2$TiSi:Co$_\text{Ti}$ system is predicted to loose
its half-metallic character.
A comparison with the low Co$_\text{Si}$-IB 
reveals qualitatively similar $\vec k$-distributions in the BSF, 
with the amplitude increased far away from the $\Gamma$-point. 
The two cases are, however, quantitatively substantially different.
As indicated by the different multiplicative factors applied,
the Co$_\text{Ti}$-BSF amplitude is about twice as large
as compared to the Co$_\text{Si}$ one. 
One further notes, in addition, the much weaker broadening
of the BSFs in this system, with the alloy bands largely
preserving their bulk character. 

The features related to the Co anti-sites 
discussed here are in very good agreement, 
both in their energy position and in 
their amplitude, to those reported in the
literature for other Co-based full Heusler alloys such as
Co$_2$MnSi,\cite{PCF04,HSK09b} Co$_2$MnGe,\cite{PCF04}
Co$_2$CrAl,\cite{AH13} and Co$_2$VSn.\cite{MKHM13}

(4) Another IB falling inside the half-metallic gap
is the one related to the Ti$_\text{Co}$ anti-site, 
shown in \FG{Co2TiSiDefBSFdn}(d). 
One notes that, compared to the previously surveyed IBs,
this has the highest intensity, no up-scaling 
having been applied in this panel. 
It is located $\simeq 0.2$~eV below $E_F$,
thus bearing no influence on the half-metallicity of
the system and, in contrast to the previous ones,
is strongly peaked around the BZ center.

(5) Finally, the signature of a Ti$_\text{Si}$-related IB
can be seen in \FG{Co2TiSiDefBSFdn}(e) 
in the vicinity of the $X$-point, at $\simeq 0.3$~eV
above the Fermi energy. It is close to
and eventually merges into a band complex 
stemming from the d-orbitals of the Ti
sublattice. Owing to its location, this IB has no direct
influence neither on the half-metallic nor on the transport 
properties of the system.

Highlighted by red ellipses in \FG{Co2TiSiDefBSFdn}
are the spin-mixing d-d band anti-crossings which
lie in the vicinity of the Fermi energy. We focus
on the minority-spin image of the falling 
majority-spin band. Its spectral function shows
variations from system to system, with the 
SOC-mediated spin-mixing effect 
strongly depending on the defect type. 
The Vc$_\text{Co}$ and Co$_\text{Si}$ defects
are characterized by a continuous BSF with 
large values extending deep into the half-metallic
gap. For the Ti-sublattice-related
defects Co$_\text{Ti}$ and Si$_\text{Ti}$,
the amplitude of spin mixing is smaller and
the BSF exhibits merely a discrete rather
than a continuous character. 
In the case of the Ti$_\text{Co}$ and Ti$_\text{Si}$ anti-sites,
the flat Co band preserves almost completely
its minority-spin character. The band repulsion is 
very weak and large BSF values are found only
in a small energy interval around the anti-crossing point.

The most important observation to be made at this point
concerns the possible changes in conductivity that may 
result from these anti-crossings. This is because
the two alloy bands involved differ not only in
their dominant spin character, but also in their
associated Fermi velocity. One can
think of a defect creating additional 
minority-spin states in the appropriate energy range. 
As a result of the spin mixing, these states will not only 
gain majority-spin character; but they will also
{\em borrow the dispersion} of the band with which they mix.
A direct consequence, shown below to occur in
the two systems containing Vc$_\text{Co}$ and 
Co$_Z$, is an increased conductivity in the
energy range near the band mixing.

We close this section with a brief comment on the dependence of
the discussed effects on the group IV element $Z$ in the
Co$_2$Ti$Z$ series. Taking as an example the Co$_Z$ anti-site system 
modeled as Co$_2$Ti($Z_{0.97}$Co$_{0.03}$), 
we show in \FG{Co2TiZCoZBSF} the BSFs of the three different alloys
for both spin components. 

The majority-spin states (top panel) 
close to the Fermi energy do not vary too
much when changing the $Z$-atom. 
A slight upward shift of the valence bands 
can be noted upon increasing the atomic
number from Si to Sn. It originates from
hybridization between the Co-d and the
$Z$-p states which occurs already in the 
ordered compounds,\cite{SSK10,PCF02}
and is manifesting itself in both spin channels.
Indeed, as seen in the bottom panel of
\FG{Co2TiZCoZBSF}, also the minority-spin d-bands
are found at higher energies in 
Co$_2$TiSn:Co$_\text{Sn}$ 
as compared to Co$_2$TiSi:Co$_\text{Si}$.
The effect is more pronounced in the 
valence band, leading to a monotonous
reduction of the half-metallic gap 
in the series Si$\to$Ge$\to$Sn. 
The two IBs related to the Co$_Z$ anti-site 
follow a similar upwards shift while maintaining
a nearly constant separation. The striking resemblance 
of the minority-spin BSFs in the Si- and Ge-based systems
should be noted here, while Co$_2$TiSn:Co$_\text{Sn}$ exhibits 
qualitative differences. On the one hand side, 
the low IB approaches the Fermi energy, thus
strongly reducing its half-metallic character.
The upper IB, on the other hand, 
moves above the Co-d band at $0.3$~eV, 
with a subsequent strong reduction of the
BSF around the $\Gamma$-point. 

\begin{figure}
  \centering
  \includegraphics[width=0.49\textwidth]{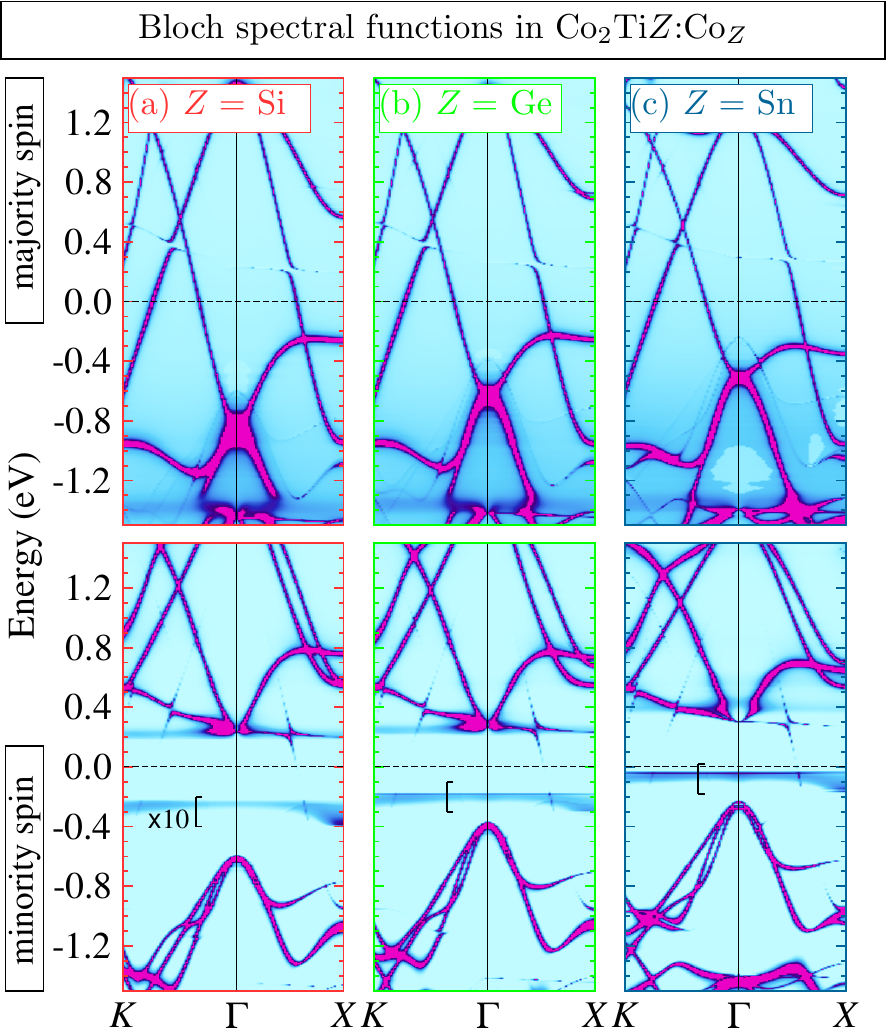}
    \caption{(Color online) Spin-resolved Bloch spectral functions 
      for the Co$_2$Ti$Z_{0.97}$Co$_{0.03}$ alloys modeling the
      Co$_Z$ anti-site, with $Z=$ Si, Ge, and Sn corresponding to the
      panels (a), (b), and (c).}
  \label{Co2TiZCoZBSF}
\end{figure}

\section{Influence of defects on the transport 
          properties}\label{SecTransport}

This section presents the calculated 
transport properties of the Co$_2$Ti$Z$:$D$ systems,
obtained by employing the CPA-based 
dilute alloy modeling.
We start with a detailed analysis of the
series of Co$_2$TiSi:$D$
systems with $3$~\% defect concentration
surveyed in the previous section.
The results obtained
for the energy-dependent electronic conductivity
are used to interpret the calculated Seebeck coefficient,
shown in \FG{SeebeckCo2TiSiDef}. 
The connection of both quantities with the
particular features of the underlying electronic structure 
will be appropriately evidenced. Our discussion will then be 
extended by covering different defect concentrations
and all the group IV elements $Z$ in Co$_2$Ti$Z$:$D$.
We show that the conclusions derived from the 
Co$_2$TiSi:$D$ systems are of general validity
in the dilute limit, with 
the Seebeck coefficient exhibiting only a weak 
dependence on both factors.

\subsection{Electronic conductivity  in Co\bm{_2}TiSi:\bm{D}}

The calculated energy dependent 
electronic conductivity of the Co$_2$TiSi:$D$
dilute alloys is shown in \FG{SigmaXXCo2TiSiDef}.
We only plot the 
longitudinal $\sigma_{xx}(E)$,
given relative to the Fermi energy $E_F$ of each system 
(dashed vertical line). 
Indeed, we found that, 
over a broad energy range, $E_F\pm 0.1$~eV, the difference
$\sigma_{zz}(E)-\sigma_{xx}(E)$,
a measure for the anisotropic magnetoresistance (AMR),
does not exceed $1$~\% for all the investigated
systems.

\begin{figure}
 \includegraphics[width=0.49\textwidth]{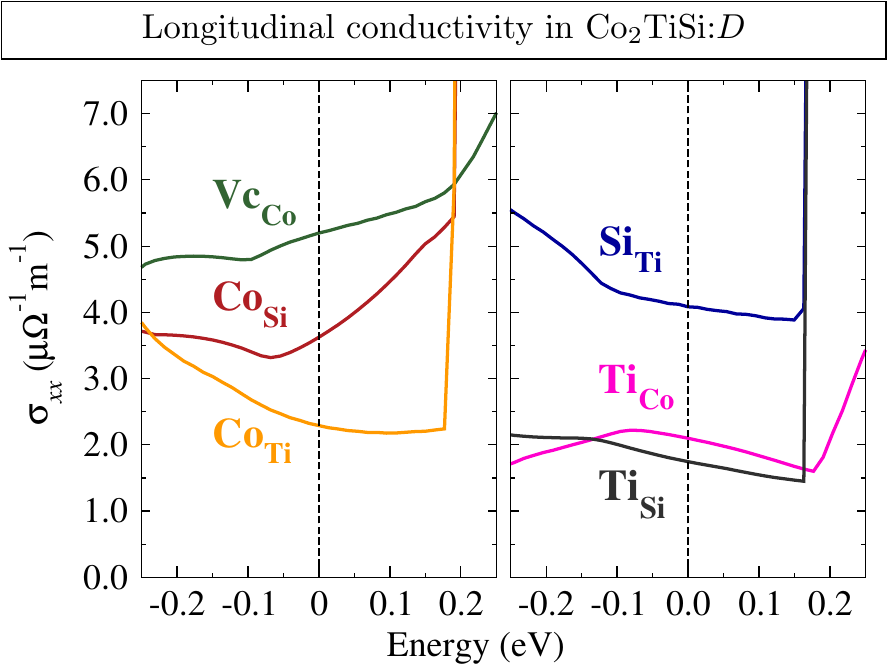}
 \caption{(Color online)
  Energy dependence of the diagonal element $\sigma_{xx}(E)$ of
  the electronic conductivity tensor, 
  calculated for the same off-stoichiometric native defects in 
  Co$_2$TiSi as those shown in \protect\FG{SeebeckCo2TiSiDef}.}
 \label{SigmaXXCo2TiSiDef}
\end{figure}

The results of our calculations indicate a strong variation 
of the residual conductivity $\sigma_{xx}(E=E_F)$ with the 
type of defect, with values ranging from $1.75$~$\mu\Omega^{-1}\text{m}^{-1}$
for Ti$_\text{Si}$ to $5.20$~$\mu\Omega^{-1}\text{m}^{-1}$ 
for Vc$_\text{Co}$ at $3$~\% defect concentration. 
All these values are higher than the
$0.45$~$\mu\Omega^{-1}\text{m}^{-1}$ deduced from the
experimentally reported residual resistivity,\cite{BFB+10}
indicating additional sources of carrier scattering in the samples.
Consistent with the decrease of $\sigma_{xx}(E=E_F)$ with defect 
composition observed in our calculations, 
these could be point defects in high concentration and/or other 
crystal imperfections. Analogous results were
obtained also for Co$_2$TiGe:$D$ and 
Co$_2$TiSn$:D$.
These defect-dependent {\em quantitative} differences directly
reflect the changes occurring in the majority-spin bands 
that cross the Fermi energy. 
Qualitatively, while a smooth dependence of all
$\sigma_{xx}(E)$-curves in
the $E_F\pm 0.15$~eV energy interval is obvious,
several peculiarities can be noted:
(i) in four cases the conductivity dependence
on its energy argument exhibits
a negative slope around $E_F$;
(ii) a steep increase of $\sigma_{xx}(E)$ occurs 
at energies above $\simeq 0.18$~eV; 
(iii) in two of the investigated systems, 
Co$_2$TiSi:Vc$_\text{Co}$ and Co$_2$TiSi:Co$_\text{Si}$, 
the conductivity is characterized by a {\em positive} 
slope around the Fermi energy. The reasons for these qualitative 
differences are discussed in detail in the following.

(i) The negative slope in the energy dependence of 
$\sigma_{ii}(E)$ around $E_F$ obtained in the case 
of Co$_\text{Ti}$, Si$_\text{Ti}$,
Ti$_\text{Co}$ and Ti$_\text{Si}$ defects is similar to that
found in the bulk Co$_2$Ti$Z$ systems.\cite{GKP14,BFB+10} 
Indeed, for all these materials the DOS
decreases above $E_F$ such that, in the approximation of
an energy-independent scattering,
the conductivity is expected to decrease with increasing
energy $E$. One can then conclude that, at
least in a range extending not far away from the Fermi
energy, the presence of these four defects leads to
a weak, nearly energy-independent scattering 
of the electrons. 
This generic behavior can be modified,
however, if IBs are present. This is the case for
the Ti$_\text{Co}$ anti-site, where
the IB found below $E_F$ [labeled (4)
in \FG{Co2TiSiDefBSFdn}(d)] adds a hole-like contribution
to the conductivity. The $\sigma_{xx}(E)$-curve for this
system (right panel of \FG{SigmaXXCo2TiSiDef})
exhibits a peak below the Fermi energy, not seen
for example, in the Ti$_\text{Si}$ case.
Let us note that no evidence
was found for contributions to $\sigma_{ii}(E)$ in
the case of Co$_\text{Si}$- or Co$_\text{Ti}$-IBs
lying below $E_F$.

(ii) The strong increase in $\sigma_{ii}(E)$ at higher 
energies, $\simeq 0.18$ above $E_F$, 
is caused by the onset of the
minority-spin conduction band stemming from the Co-d orbitals.
The rise to a plateau of $12-14$~$\mu\Omega^{-1}\text{m}^{-1}$
appears as an abrupt jump due to the chosen scale of the figure. 
One has to note here that we have purposefully chosen a narrow 
interval for displaying $\sigma_{xx}(E)$ in order to emphasize 
on its $E$-dependence around $E_F$. 
We have thoroughly checked the results against numerical
in the $E$-range around the raising point 
by refining the BZ-integration for a $\vec k$-mesh as high
as $6\cdot10^6$ points and found no smoothening 
in $\sigma_{xx}(E)$. Moreover, the calculated $\sigma_{xx}(E)$ values
correspond to $T=0$~K, such that, at any finite temperature, 
the sharp steps will be washed out. For two of the
systems, containing the Vc$_\text{Co}$- and Ti$_\text{Co}$-defects,
the $\sigma_{xx}(E)$ increase is less abrupt.
In these cases the minority-spin Co band, stemming
from the perturbed sublattice, 
is effectively broadened as a result of alloying. 
Moreover, the presence of the impurity 
on a single Co sublattice [see \FG{FigStruc}(b)]
destroys the inversion symmetry of the host crystal.
Under this condition, SOC leads to scattering
between states that are no longer degenerate,
increasing the magnitude of the 
energy-dependent scattering.

(iii) In contrast to the situation above, the positive slope
of $\sigma_{xx}(E)$ near $E_F$ observed for
the Co$_2$TiSi:Vc$_\text{Co}$ and Co$_2$TiSi:Co$_\text{Si}$
indicates a strong energy-dependent scattering
occurring in these systems.
Following the inversion symmetry suppression 
in the former and the presence of a strong impurity band 
in the latter, the electronic conductivity 
increases right above $E_F$. In both cases
we suggest that a SOC-mediated broadening and
redistribution of states taking place
around the anti-crossing points is responsible
for the observed effect. As pointed out above
in the BSF analysis, additionally created 
minority-spin states may borrow mobility, via
spin-mixing, from the majority-spin bands.

\subsection{Seebeck coefficient  in Co\bm{_2}TiSi:\bm{D}}

By virtue of \EQs{Lmoms} and (\ref{EQSeb}), the 
Seebeck coefficient $S(T)$ is expressed as the quotient
of the first and zeroth moments of the conductivity.
The findings related to the changes in the 
energy dependence of $\sigma_{ii}(E)$
can then be directly transferred to the 
corresponding results obtained for $S(T)$
presented above in \FG{SeebeckCo2TiSiDef}. 
One can easily understand the broad range of values, varying
both in sign and magnitude, obtained for the different
kinds of defects as arising from 
variations in either ${L}^{(0)}_{ii}$ (denominator) 
or ${L}^{(1)}_{ii}$ (numerator).

Assuming an energy-independent scattering, our theoretical
results predict a positive Seebeck coefficient for the
bulk Co$_2$Ti$Z$ materials,\cite{GKP14,BFB+10} 
a finding which is in contrast to the large, negative
values reported experimentally, 
reaching as much as $-30$~$\mu$V/K in 
Co$_2$TiSi/Ge and $-50$~$\mu$V/K in
Co$_2$TiSn.\cite{BFB+10,BOG+10} 

As seen in \FG{SeebeckCo2TiSiDef}, this tendency of
a positive Seebeck coefficient is retained for the 
systems that have a weakly energy-dependent scattering,
such as Co$_\text{Ti}$, Ti$_\text{Co}$, Si$_\text{Ti}$, 
and Ti$_\text{Si}$. As the temperature increases,
the contribution of the minority-spin electrons
becomes more important. This is because the minority-spin bands 
above the half-metallic gap become populated at higher temperatures.
As shown above, in the case of defects with a weakly
energy-dependent scattering the onset of conductivity 
in the minority-spin bands leads to an abrupt rise in the conductivity 
and thus to a negative, electron-like contribution to the Seebeck 
coefficient. As a result the Seebeck drops sharply and becomes 
negative at elevated temperatures for the systems
containing Co$_\text{Ti}$, Si$_\text{Ti}$, and Ti$_\text{Si}$.
One notes here the direct relation between the 
change of slope for $S(T)$ and the ever increasing
contribution coming from the flat minority-spin band.
The exception to this behavior is represented
by the Ti$_\text{Co}$ anti-site where the presence
of the IB below $E_F$ was shown to add
a hole-like contribution to the conductivity,
a contribution which is large enough such as to
preserve a positive Seebeck coefficient throughout the whole
investigated temperature range.

In the systems with a strong energy-dependent
scattering, the conductivity was found to be 
larger above $E_F$ than below. This leads to a negative 
Seebeck coefficient, as obtained for the Co$_2$TiSi:Vc$_\text{Co}$ 
and Co$_2$TiSi:Co$_\text{Si}$.
We note that only in these cases the results 
qualitatively reproduce the experimentally 
observed behavior. For the Co$_\text{Si}$ anti-site
the calculated $S(T)$ deviates within $30$~\%
from the experimental data for $T\leq 300$~K. 
As we will show below, only varying the defect
concentration is not sufficient to further improve the 
agreement between our calculations and experiment, 
which is particularly deteriorating at higher
temperatures.
Indeed, it is expected that spin fluctuations
(not accounted for here) become important when the Curie 
temperature is approached, and that 
magnon drag effects will lead to an enhanced thermopower.
We note that, within a hydrodynamic theory of such effects, 
the sign of the Seebeck coefficient obtained from an 
independent-electron calculation is preserved.\cite{WDT+16}

\subsection{Effect of defect concentration and of group 
               IV element}

We have established so far the needed link between the
electronic structure and, through the 
energy dependence of the electronic conductivity,
the main features of the Seebeck
coefficient in the Co$_2$TiSi:$D$ systems.
We now extend our analysis to a broader quantitative level
by investigating two effects: (i) a varying defect concentration,
and (ii) different group IV element in the Co$_2$Ti$Z$:$D$ series.

\begin{figure}
 \includegraphics[width=0.49\textwidth]{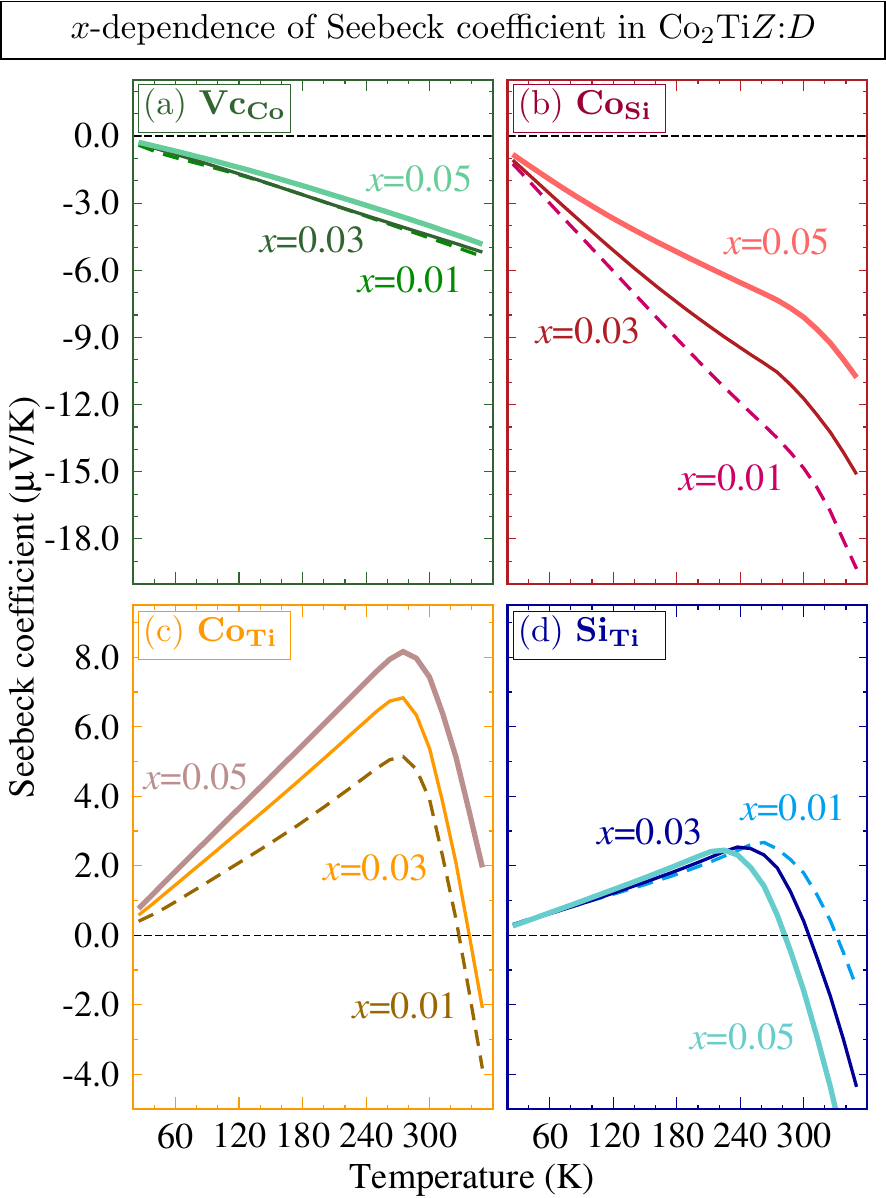}
 \caption{(Color online) 
   Seebeck coefficient $S(T)$ calculated for
   several off-stoichiometric native defects in 
   Co$_2$TiSi:$D$. With each defect $D_A$ 
   modeled as a single-site effective 
   medium $A_{1-x}D_x$, the dashed, thin solid, and thick solid lines
   correspond to $x=0.01$, $x=0.03$, and $x=0.05$ respectively.
   Panels in the same row have identical scales for $S(T)$.}
  \label{SEBCo2TiSiDefVarX}
\end{figure}

Figure~\ref{SEBCo2TiSiDefVarX} shows the calculated 
temperature-dependent Seebeck coefficient $S(T)$ for
three different off-stoichiometric compositions
$x=0.01$, $x=0.03$, and $x=0.05$. 
Due to their fairly similar behavior to that of the 
Co$_\text{Ti}$-system, the corresponding Ti anti-sites
related curves Ti$_\text{Co}$ and Ti$_\text{Si}$ have been
omitted in this figure.

We start by noting that a monotonous decrease of 
$\sigma_{ii}(E)$ as a function of defect composition 
was observed for all investigated systems.
Accordingly reflected in the integrated quantities
defined by \EQ{Lmoms}, $L_{ii}^{(0)}$ decreases with $x$
irrespective of the defect considered.
The differences in the $S(T)$ 
concentration dependence noticed
in \FG{SEBCo2TiSiDefVarX}, weak for 
Vc$_\text{Co}$ and Si$_\text{Ti}$ and rather pronounced
for Co$_\text{Ti}$ and Co$_\text{Si}$, can be related
to the behavior of the $L_{ii}^{(1)}$ term, 
measuring the {\em asymmetry} of $\sigma_{ii}(E)$ about the 
Fermi energy. 

Indeed, if this term remained constant with  $x$, one would obtain 
a monotonous increase in the absolute value of the Seebeck coefficient,
$|S(T)|$, which is obviously the case for the Co$_\text{Ti}$ defect, 
panel (c) of \FG{SEBCo2TiSiDefVarX}.
For Vc$_\text{Co}$ and Si$_\text{Ti}$ [panels (a) and (d)]
the nearly concentration-independent $S(T)$ indicates an 
asymmetry term that decreases with $x$, without changing
its sign, thus compensating the decrease of $L_{ii}^{(0)}$. 
Note that, in the case of Si$_\text{Ti}$, the shift of the 
$S(T)$ peak is caused by a slight upward shift in the Fermi 
energy taking place with increasing Si content.
An even faster decrease of $L_{ii}^{(1)}$ was found to occur
for the Co$_\text{Si}$ system [panel (c)]. All these
cases can be understood as a consequence of a
broadening and smoothening, with increasing concentration $x$,
of the additional contributions to $\sigma_{ii}(E)$
from above $E_F$.

Analogous considerations can be applied when analyzing the
dependence of the Seebeck coefficient on the group IV
element $Z$ in the isoelectronic systems Co$_2$Ti$Z$:$D$.
The corresponding results for all investigated defects
at a net composition of $x=0.03$ are shown in \FG{SEBCo2TiZDef}.

\begin{figure}
 \includegraphics[width=0.49\textwidth]{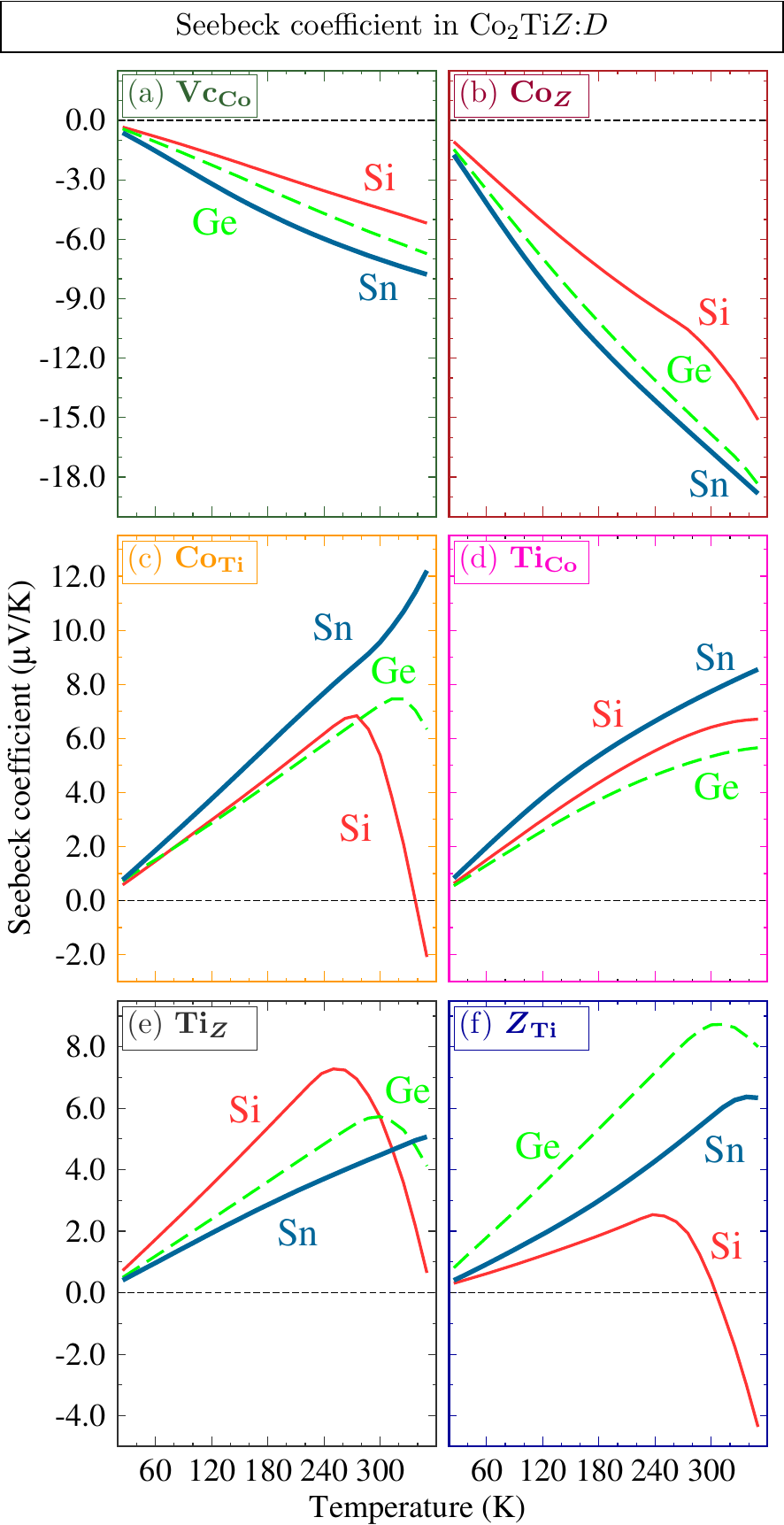}
 \caption{(Color online) 
   Seebeck coefficient $S(T)$ calculated for
   the selected defects in Co$_2$Ti$Z$:$D$ at
   $3$~\% defect composition. In each panel,
   the curves corresponding to different group IV elements $Z$ are
   represented by the (red) thin solid, (green) long dashed, 
   and (blue) thick solid lines for Si, Ge, and Sn, respectively.
   Panels in the same row have identical scales for $S(T)$.}
 \label{SEBCo2TiZDef}
\end{figure}

Although the conductivity is mainly dominated by 
majority-spin carriers, one can ascribe the 
various differences in the $Z$-atom dependence of $S(T)$ 
as being mostly related to changes occurring in the 
minority-spin channel. Indeed, the BSFs displayed in
\FG{Co2TiZCoZBSF} for the Co$_2$Ti$Z$:Co$_\text{Si}$ system 
hint to a nearly unchanged majority-spin DOS around $E_F$ 
for all three systems with $Z=$ Si, Ge, and Sn. 
Yet there is no simple rule, applicable to all defects,
that would explain how $\sigma_{ii}(E)$ depends on the
atom $Z$ of the host material. The lack of such a rule
is reflected in the Seebeck coefficient being
the largest in Co$_2$TiSi:Ti$_\text{Si}$ for the Ti$_Z$ anti-site
[panel (e) of \FG{SEBCo2TiZDef}] and the smallest in 
Co$_2$TiSi:Si$_\text{Ti}$ for $Z_\text{Ti}$ [panel (f)].

We have previously identified three systems,
Co$_\text{Ti}$, Ti$_Z$, and $Z_\text{Ti}$, 
which are characterized by a sudden change of slope in $S(T)$ at
elevated temperatures. 
This behavior was associated with
the minority-spin bands above the half-metallic gap 
becoming populated with increasing temperature. As the
offset of these bands relative to $E_F$ increases with 
$Z$ (see \FG{Co2TiZCoZBSF}), the peak in $S(T)$ 
also shifts towards higher $T$-values when $Z$ changes
from Si to Ge and then Sn. 

The absence of such a peak in Co$_2$TiSi:Ti$_\text{Co}$ was shown to
be caused by the IB present $0.1$~eV below $E_F$, states introducing a
rather large hole-like contribution to the conduction. A similar IB is
present in all Co$_2$Ti$Z$:Ti$_\text{Co}$ systems, with its relative
position inside the half-metallic gap remaining fairly the same.

For the two defects for which the Seebeck coefficient
had a negative sign, Co$_2$TiSi:Vc$_\text{Co}$ and 
Co$_2$TiSi:Co$_\text{Si}$, we find this sign to persist
for Co$_2$TiGe and Co$_2$TiSn.
Moreover, we note that for both defects $|S(T)|$ {\em increases} 
with increasing atomic number of the group IV element $Z$,
a rather non-intuitive behavior. 
Indeed, the calculated $\sigma_{ii}(E)$, 
which is governing the size of $L_{ii}^{(0)}$, also increases 
monotonously along the series Si$\to$Ge$\to$Sn. 
It follows that the increase in $|S(T)|$ can only be accounted for 
by a much larger increase in $L_{ii}^{(1)}$, the asymmetry term.
Noting that the SOC strength  also 
increases as moving down in the group IV, we conclude
that the obtained results are consistent with the 
interpretation given about the role of the SOC in 
causing a largely asymmetric conductivity.
It consists of a redistribution of states in the
systems lacking inversion symmetry (as in
Co$_2$Ti$Z$:Vc$_\text{Co}$) and an increased
mobility of the minority-spin IB by mixing it
with the majority-spin band crossing the Fermi energy,
in Co$_2$Ti$Z$:Co$_Z$.

\section{Conclusions}

The Ti-based Heusler alloys Co$_2$Ti$Z$ ($Z=$ Si, Ge, Sn) have been
studied by means of density-functional calculations with the goal to
explore their suitability for spintronics and spincaloric
applications. Since the ideal Heusler alloys of this composition are
ferromagnetic half metals, it is of particular interest to learn how
the half-metallicity is modified in realistic materials, including
off-stoichiometry and atomic-scale disorder. To this end, the formation
energies $E_\text{form}$ of intrinsic point defects have been
calculated, and their consequences for electronic transport and
thermopower have been explored.  

While the formation of anti-structure defect pairs by atomic swaps is
generally found to be energetically costly, our calculations indicate
a very high sensitivity of the ternary Heusler alloys to deviations
from their ideal stoichiometry. Once formed in appreciable numbers,
point defects caused by off-stoichiometry significantly influence the
electrical conductivity. Moreover, the Seebeck coefficient turns out
to be  a sensitive probe for the presence of defects,
even in the simple situation treated here, of
single defect formation, rather than interacting pairs.

The lowest values of E$_{\rm form}$ are found  for Vc$_\text{Co}$ and
Ti$_Z$, followed by the Co-related anti-sites Co$_\text{Ti}$ and
Co$_Z$.  The formation energy of Ti$_Z$ may become negative
in the case of $Z=$ Sn, and is still small in the Si and Ge compounds
($0.48$~eV and $0.30$~eV, respectively). 
Under Co-deficient conditions, the formation energy of 
Vc$_\text{Co}$ may become negative
for all three alloys studied, 
indicating that Co vacancies may occur spontaneously.
Intrinsic point defects are to be
expected also under Co-rich conditions, 
since E$_\text{form}$ of Co$_Z$ and of Co$_\text{Ti}$ are
fairly low (in the $0.3$ to $0.6$~eV range) for Heusler alloys containing
Si or Ge, while being somewhat higher in Co$_2$TiSn. Hence,
anything that comes close to
ideal Heusler alloys can be synthesized only under very special
conditions when the stoichiometry is met exactly, since slight
deviations from stoichiometry, both to the Co-rich or the Co-poor
side, are likely to lead to point defect formation. 
Further support for the abundance of defects  comes from measurements 
of the electrical conductivity, where a high residual resistivity, 
and only a moderate increase (in the range of $30$ to $50$~\%) 
from low- to room-temperature, have been
observed experimentally.\cite{BFB+10} This points to a relatively high 
concentration of native defects in the material, formed already
during sample preparation. 

While one would expect from the density of states at the Fermi energy
a positive Seebeck coefficient in all three alloys, experiments
reported negative values throughout the whole
temperature range.\cite{BFB+10,BOG+10} 
In this context, it is important that our calculations of electronic
transport demonstrate a
strong effect of point defects on the thermopower of the Co$_2$Ti$Z$
Heusler alloys. At defect concentrations of a few percent, 
the results obtained here indicate that the magnetic order and the 
half-metallicity in these materials is largely retained. Hence, 
the majority-spin carriers dominate the electronic transport, 
ruling out earlier explanations of the negative Seebeck coefficient 
being due to a breakdown of magnetism and a non-magnetic electronic 
structure. Instead, we find that the negative Seebeck coefficient 
is due to subtle defect-induced changes of the electronic structure that
occur already for small concentrations of  Co$_Z$ and, 
to a lesser extent, of Vc$_\text{Co}$ defects. Both defect types are 
likely to occur in Co-rich and Co-poor samples, respectively: 
While E$_{\rm form}$ of Vc$_\text{Co}$ may even be negative,  
Co$_Z$ has a formation energy of only $0.30$~eV in Co$_2$TiSi and 
$0.45$~eV in Co$_2$TiGe, being second lowest after Vc$_{\rm Co}$. 
Hence, both for surplus or a deficiency of Co, the
defects with the lowest formation energy induce a negative
thermovoltage. Thus, the results of our calculations help to
rationalize  the experimental findings at low temperatures. In
addition, we point out that, both in ideal Ti-derived Heusler alloys
as well as in many defected materials (Co$_\text{Ti}$, Ti$_Z$,
$Z_\text{Ti}$), the minority-spin electrons start to contribute 
to the conductivity, and hence also to the Seebeck coefficient. 
This is another factor responsible for a persistently negative 
Seebeck at higher temperatures.

With the Curie temperature of the Co$_2$Ti$Z$ lying in the
range of $380$-$400$~K,\cite{BFB+10}
spin fluctuations are expected to start to play a role 
at room temperature and above. Via the pertinent 
scattering mechanisms for the charge carriers, 
they might also have an impact on the Seebeck coefficient in 
this temperature range. Investigations of such complex effects,
also including the temperature-induced lattice vibrations
as well as a detailed analysis on the spin-dependence of
the Seebeck coefficient will make the subject 
of future work. 

In summary, the presented computational results highlight the decisive
role played by off-stoichiometry in Heusler compounds, which must be
taken into account when one wishes to  select materials from this
class for specific applications. The thermopower is found to be a
sensitive probe for the existence of defects, but remains difficult to
interpret even with computational results at hand, in particular in
cases where several contributing factors | different types of point
defects, phonon and magnon scattering | are superimposed. 


\acknowledgments

This work was supported by the German Research Foundation 
({\em Deutsche Forschungsgemeinschaft -- DFG}) 
within the Priority Program 1538 ''Spin Caloric Transport
(SpinCaT)''. 
The authors acknowledge the computing time granted by the 
John von Neumann Institute for Computing (NIC) and provided on the 
supercomputer JURECA at J\"ulich Supercomputing Center (JSC).
Additional access to the Cray-XT6m supercomputer facility
has been offered by the Center for Computational Sciences 
and Simulation (CCSS) at the University Duisburg-Essen. 


\begin{thebibliography}{63}
\bibitem[{FPv()}]{FPvsASAnote}
\bibinfo{note}{A side-by-side comparison of ASA vs.\ FP density of states
  curves in Co$_2$TiSn is shown in Fig.~5 of Ref.~\onlinecite{KKW+07}. While
  the FP approaches predict this compound as being half-metallic, it is
  metallic in the ASA-based calculations.}
\bibitem[{Seb()}]{SebNote}
\bibinfo{note}{Here the off-diagonal elements of the tensors
  $\underline{L}^{(0)}$ and $\underline{L}^{(1)}$ connected to the anomalous
  Hall and Nernst effects were neglected. In cubic metals this assumption is in
  general well-justified by the significantly smaller magnitude of these purely
  antisymmetric tensor elements in comparison to the diagonal ones.}
\bibitem[{Cur()}]{CurtaNote}
\bibinfo{note}{In Ref.~\onlinecite{CC09}, the Co$_2$Ti most stable crystal
  structure is predicted to be $C14$, with a formation energy lying
  $0.022$~eV/f.u. below that of $C15$.}
\end{thebibliography}
\end{document}